\documentclass[a4paper,12pt,bezier,fullpage]{article}
\usepackage{mathptmx}
\usepackage{epsf}
\usepackage[T1]{fontenc}
\usepackage{latexsym}
\usepackage[utf8]{inputenc}
\usepackage{gensymb}
\usepackage{graphicx}
\usepackage[a4paper, margin=1.5cm]{geometry}
\usepackage{amsmath}
\usepackage[nottoc,numbib]{tocbibind}
\usepackage{hyperref}
\usepackage{bigints}
 \usepackage{xcolor,colortbl}
 \usepackage{float}
 \usepackage{authblk}
 \usepackage{siunitx}
 \usepackage{setspace}

 \newcommand{\beginsupplement}{%
        \setcounter{table}{0}
        \renewcommand{\thetable}{S\arabic{table}}%
        \setcounter{figure}{0}
        \renewcommand{\thefigure}{S\arabic{figure}}%
        \renewcommand{\thesubsection}{S.\Roman{subsection}}%
        \renewcommand{\thesubsubsection}{\roman{subsubsection}}%
     }
     
\usepackage{multicol}
 \usepackage[font=footnotesize,labelfont=bf,justification=justified,format=plain]{caption}

\title{\emph{Stereo Darkfield Interferometry} : a versatile localization method for subnanometer force spectroscopy of single molecules and 3D-tracking of single cells.}
\author[1,*]{Martin Rieu}
\author[1,*,3]{Thibault Vieille}
\author[1,3]{Gaël Radou}
\author[1]{Raphaël Jeanneret}
\author[2]{Nadia Ruiz}
\author[1]{Bertrand Ducos}
\author[1]{Jean-François Allemand}
\author[1]{Vincent Croquette}
\affil[*]{These authors contributed equally to this work}
\affil[1]{Laboratoire de physique de L’École normale supérieure de Paris, CNRS, ENS, Université PSL, Sorbonne Université, Université de Paris, 75005 Paris, France}
\affil[2]{Institut de Biologie de l’Ecole Normale Supérieure (IBENS), Ecole normale supérieure, CNRS, INSERM, Université PSL, Paris, France}
\footnotetext[3]{Present address :  $\text{Depixus}^\text{\textregistered}$ 3-5 Impasse Reille, 75014 Paris}
\begin{document}

\maketitle

\begin{abstract}
        Super-resolutive 3D tracking, such as PSF engineering or evanescent field imaging has long been used to track microparticles and to enhance the throughput of single molecules force spectroscopic measurements. However, current methods present two drawbacks. First, they lack precision compared with optical tweezers or AFM. Second, the dependence of their signal upon the position is complex creating the need for a time-consuming calibration step.

Here, we introduce a new optical technique that circumvents both issues and allows for a simple, versatile and efficient 3D tracking of diluted particles while offering a sub-nanometer frame-to-frame precision in all three spatial directions. The principle is to combine stereoscopy and interferometry, such that the z (axial) position is measured through the distance between two interferometric fringe patterns. The linearity of this stereoscopy technique alleviates the need for lookup tables while the structured interferometric pattern enhances precision. On the other hand, the extended spatial footprint of this PSF maximizes the number of photons detected per frame without the need of fancy cameras, nor the need for complex hardware. Hence, thanks to its simplicity and versatility, we believe that SDI (Stereo Darkfield Interferometry) technology has the potential to significantly enhance the spreading of 3D tracking.

We demonstrate the efficiency of this technique on various single-molecule measurements thanks to magnetic tweezers. In particular we demonstrate  the precise quantification of two-state dynamics involving axial steps as short as 1 nm. We then show that \emph{SDI} can be directly embedded in a commercial objective providing a means to track multiple single cells in 3D .

\end{abstract}
\begin{multicols}{2}
\section{Introduction}

Video-based 3D tracking of micro-particles has become a major tool revealing the dynamics of microorganisms motility \cite{berg1972chemotaxis} and the kinetics of molecular motors interacting with biopolymers tethered to microbeads \cite{hodeib2016single}.  However, in spite of recent significant improvements  \cite{brouwer2020multiplexed} \cite{dulin_high_2015} \cite{huhle_camera-based_2015}, these techniques suffer from the limitation of their precision and from the requirement of object-specific calibrations prior to any measurement. In particular, while these camera-based methods present the extremely useful advantage of being parallelizable,  they were not able to reproduce the very precise nanometric measurements made with the optical tweezers (OT) or AFM  that revolutionized the understanding of molecular motors stepping \cite{ragan_3d_2006}\cite{yildiz_kinesin_2004} \cite{abbondanzieri_direct_2005}\cite{greenleaf_direct_2008}. The reason behind this difference lies in the huge number of photons recorded in OT or AFM in comparison with the one obtained using camera sensors.

The last decade has witnessed significant improvements \cite{von_diezmann_three-dimensional_2017} of microparticles 3D tracking techniques, providing better axial and lateral precision and smaller optical aberrations. However, most of these competing solutions have made improvements solely for one feature. For instance, while some allow subnanometer precision, they are not suited for parallelization \cite{abbondanzieri_direct_2005}. Others do not allow long real-time tracking,  as they rely on very high acquisition rates \cite{kim_high-resolution_2009} incompatible with the maximal transfer rate between cameras and computers.

Among these methods, evanescent fields based ones are precise but limited in their detection range, of the order of the wavelength of the exciting light \cite{hanne_opening_2007}\cite{brutzer_scanning_2012}\cite{graves_dynamic_2015}. Technologies based on the dependence of the point spread function (PSF) with the axial position \cite{taute_high-throughput_2015} \cite{gosse_magnetic_2002} \cite{baddeley_three-dimensional_2011} \cite{shechtman_optimal_2014} \cite{kao_tracking_1994} \cite{huang_three-dimensional_2008} \cite{pavani_three-dimensional_2009} \cite{jia_isotropic_2014} allow precise 3D-tracking in the far field but suffer from the complexity of their experimental implementation and from their dependence upon the quantitative measurement of the image shape that increases their sensibility to optical imperfections such as spherical aberration or astigmatism. More importantly, the need for lookup tables calibrated for each individual emitter (or at least for sufficient positions in the field of view) \cite{von_diezmann_three-dimensional_2017} limits their throughput and their usage. Tracking of free particles variable in size quickly passing in the field of view is then almost impossible. A significant amelioration of ring-shaped PSF's  \cite{gosse_magnetic_2002}\cite{taute_high-throughput_2015} was recently proposed \cite{taylor2019interferometric}, allowing nano-particles tracking in living cells, but is limited in axial range (a few hundreds of nanometers) and precision (5-6 nm).

By comparison, stereoscopic methods \cite{sun_parallax:_2009} \cite{sun_single-molecule_2010} \cite{sancataldo2017three} are more versatile and proficient in the context of heterogeneous and fast-moving particles. Using two light sources with different incidences, they produce two images coinciding when the particle is in focus. As the tracked object moves out of focus by dz, the shift between both images dx increases linearly with dz. This linearity is a great feature of the method alleviating the requirement of a calibration step and insuring its independence with respect to the particle size. However, the necessary reduction of the numerical aperture to at least half of the possible angles results in a larger point-spread function (PSF) and thus a reduced precision. This loss of information can be in principle compensated by increasing the number of photons used to build their PSF, but the finite well-depth (the maximum number of photons treated by one pixel during one frame) of cameras sets a limit to this strategy. At the same time, such a numerical aperture limits the axial measurement range. 

Here, we present a method, \emph{Stereo Darkfield Interfetrometry} (SDI)  \cite{brevet} that combines the linearity and parallelization ability of stereoscopic methods with the high per-photon information content of structured PSF. It thus allows to optimize the use of the transmission rate and of the well depth of cameras. Furthermore, the technique enables to reach a better compromise between the axial measurement range and the resolution by selecting a small range of incident angles and by using interferometry to structure light and thus to boost precision. Our approach is designed to track in real time a large number of diluted microparticles with subnanometer axial and lateral precision with no limit on the duration of the experiments, thus enhancing significantly the throughput of subnanometer 3D single-particle tracking. Furthermore, it allows to greatly simplify the tracking of free microorganisms by suppressing the need to constitute lookup tables.

We illustrate the precision of the method by acquiring, for the first time with a camera-based method, traces from force spectroscopic events showing nanometric stepping (helicase stepping, oligonucleotide hybridization), thus paving the way for the multiplexing of measurements that were so far exclusively feasible with optical tweezers or atomic force microscopes (AFM), and for this reason hardly parallelizable.

We then illustrate the versatility of the method by measuring dynamic properties of unicellular algae through the acquisition of their 3D trajectories, thus expanding the utility of our approach to a wide range of applications in biological imaging.

\section{Setup}

\begin{figure*}[!ht]
\centering
\includegraphics[width=18cm]{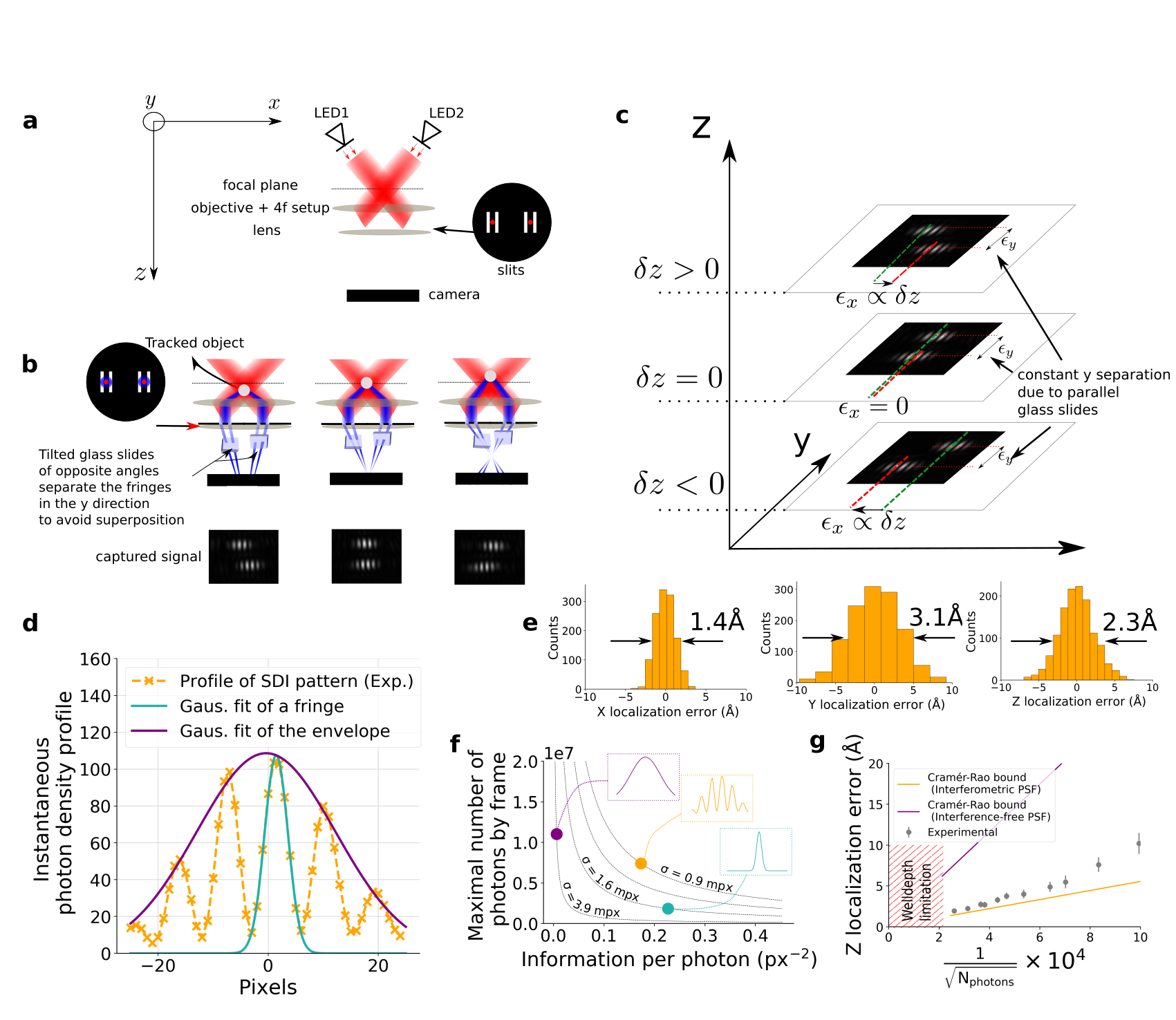} 
\caption{A schematic description of the SDI setup. The red and blue colors denote respectively parallel incident light and light diffused by the scatterer (same wavelengths). \textbf{a.} In the absence of the scatterer, the incoming parallel light is blocked, ensuring darkfield. \textbf{b.} When a scatterer is present, the light goes through the slits and creates a PSF consisting of two interference patterns. \textbf{c.} Vertical stack of the SDI images (objective 100X) as a function of the defocus (axial position $z$) of the tracked object. The transverse distance $\epsilon_x$ between the two spots is proportional to the defocus. \textbf{d.} Typical transverse density profile of a \emph{SDI} pattern. One fringe and the envelope are being fitted by Gaussians. \textbf{e.} Distribution of the inferred position, at maximal light intensity, of the 3D position of stuck microspheres. 1280 frames are analyzed. No averaging is performed. The mechanical and thermal drift are subtracted in order to assess the optical noise of the setup (see Materials and Methods). \textbf{f.} Theoretical number of photons by frame and information per photon for each profile shown in d.  Interferences allow to increase the number of photons while keeping a good precision. Thin black lines join points with equal theoretical precision $\sigma$ (values in millipixels). \textbf{g.} Standard deviation of the measured axial position of a microsphere as a function of light intensity (objective 100X). This is compared with the theoretical Cramér-Rao bounds computed from the experimental profile drawn in $d$ and its envelope. The maximal light intensity is constrained by the camera's well-depth (here 30.000 electrons per pixel).}
\label{Setup}
\end{figure*}

Figure~\ref{Setup} presents the principle of Stereo Darkfield Interferometry (\emph{SDI}). Illumination is generated by two superluminescent LEDs (SLED) sending collimated parallel beams with symmetric incidence angles of $\pm 23 \degree$ from the optical axis (Figure~\ref{Setup}.a). The microscope is built around an infinitely corrected objective and a field lens producing an image on a camera chip. An afocal system built with 4 lenses (Figure~\ref{4f}) is used to access the image focal plane of the objective that is physically located inside the objective mount.
%If you address biologists you need to explain a bit that a lens can be used to perform a Fourier transform, the Transform filtered and Fourier back (and cite references where this is detailed). 
 A black absorbant film with slits is used as a filter in the Fourier plane, \emph{i.e.} the back focal plane.  In the absence of diffusing particles, no light reaches the camera as the two illumination beams are focused on the black region located between the slits. The Fourier filter is built with four symmetrical slits spanning the $x$ axis. When a micron size particle is placed in the objective focal plane, the light stemming from each of the sources is diffused in a cone which covers mostly the two closest slits. The slits select two beams that are focused on the camera chip by a  lens placed in contact with the slits (Figure~\ref{Setup}.b). When the particle is at the focal plane, the beams originating from both slit pairs coincide. In order to separate them along the $y$ direction, two glass slides tilted with opposite incidences ($\pm 17^{\circ}$) are placed just behind each slit pairs in close contact with the imaging lens. Their incidence is chosen to be large enough so that the two images coming from both pairs of slits do not coincide. Due to the double-slit configuration, each of the beam pairs issued from the same light source, as in the so-called Young double-slit experiment, gives rise to an interferometric pattern. When the object moves away from the focal plane by $\delta z$, both patterns are translated in a shear mode with opposite directions along the $x$-axis (Figure~\ref{Setup}.c). The distance between each of them is linearly related to the axial displacement $\delta z$ (Figure~\ref{40_lin_whole}), whereas the average of their positions is related to $x$-$y$ movement of the tracked object, as in \cite{sun_parallax:_2009}.

The structure of the interferometric fringes of \emph{SDI} increases both the axial and in-plane resolution of the bead position. Indeed, assuming a photon limited noise, the optical noise $\sigma_i$ on the 3D localization of the bead along the $i$ direction must verify (see Supp. Materials \ref{thprecision}):
\begin{equation}
\sigma_i^2 > \frac{1}{g_i^2 N} \frac{1}{\bigintss\frac{f'(i)^2}{f(i)} di}
\label{cramerrao}
\end{equation} ,
where $N$ is the number of photons received during one frame, $f$ is the profile of the interferometric pattern and $g$ is the coefficient relating the signal displacement to the displacement of the tracked object. For the transverse displacement, $g_x$ is simply the geometric magnification of the optical system. The axial displacement $g_z$ depends on the magnification and on the angle of the incident light as described by the formula given in Supp. Materials \ref{grandissement}.

%If $f$ is a Gaussian of variance $w$, the latter formula is in agreement with the Thomson-Webb formula  \cite{thompson_precise_2002} :
%\begin{equation}
%\sigma^2 > \frac{w^2}{g_x^2 N}
%\label{Thomson_Webb}
%\end{equation}

%In practice, for real-time tracking, the number of photons received per seconds is limited by the transfer rate of the camera and its well-depth (maximum number of electrons per pixel).  At fixed well-depth, the maximum number of photons $N$ that can fit into a Gaussian PSF with a width $w$ evolves as $w$. Thus, given the above equation, $\sigma$ evolves as $\sqrt{w}$ : thus, spreading the PSF (for example by reducing the numerical aperture of the objective), while increasing the maximum number of photons collected in one frame by increasing illumination, would still decrease the localization precision.
 
In practice, for real-time tracking, the number of photons received per seconds is limited by the transfer rate of the camera and its well-depth (maximum number of electrons per pixel). Increasing the spread of a Gaussian PSF results in a loss of precision. However, $SDI$ allows spreading the PSF on several pixels without losing precision (Figure~\ref{Setup}.d-f) :  the PSF is indeed structured, increasing its spatial derivative and thus the value of the integral in the denominator in equation~\ref{cramerrao}. We show in Figure~\ref{Setup}.g the experimental and theoretical values of the precision of $SDI$ : the structuration of the signal allows a 4-fold improvement of the spatial precision compared to the equivalent unstructured PSF. At the maximum light intensity allowed by the camera well depth, the  frame-to-frame error on the position of the tracked objects due to photon noise is as low as 0.2 nm in $y$ and $z$ directions, and as 0.1 nm in the x direction (Figure~\ref{Setup}.e). The difference between $z$ and $x$-axis is due to a difference between the transverse and the axial magnification of the setup (see ~\ref{grandissement}). The slightly poorer precision in $y$ is due to the asymmetry of the SDI profile pattern (no structure in the $y$ direction of the slits presented in this paper).

The double-slit configuration creates a pattern that is localized in the space of spatial frequencies and that can be easily filtered out using Fourier transform and deconvolved from low-frequency signals (see Supp. Materials~\ref{algotrack}). That, combined with the reduced range of angles selected by the slits, reduces the sensitivity to overall PSF deformations and allows increased depth of field compared to traditional stereoscopy \cite{sun_parallax:_2009}. Thus, the z axial position can be tracked up to $6 \mu m$ with 40X and 100X objectives (Fig~\ref{40_lin_whole}).

Most interestingly, \emph{Stereo Darkfield Intereferometry} can be directly embedded into an objective and thus can be commercially distributed as a 3D-tracking module for any preexisting microscope. Due to the technical complexity of planar high magnification objectives, such an embedding would need to be performed by the objective supplier. However, in order to demonstrate this possibility, we disassembled a simpler X20 objective and added the SDI slits directly in the back focal plane of the objective. We brought two prisms of opposite angles directly in contact with the slits (Figure~\ref{Dessin3}a-c). These prisms produce a phase shift in the Fourier space that translates into a position shift in the real image space, preventing the superimposition of the two interferometric patterns. They play the same role as the parallel slides presented above.  In this low magnification configuration, the precision is micrometric, while the axial depth of view is greatly enhanced with a conserved linear response : the distance between the interferometric fringes depends linearly on the $z$ position over ranges of the order of 150 $\mu$m (Figure~\ref{Dessin3}.h). Residues to linearity over such a range lay below $0.8 \mu m$ and are shown in Figure~\ref{residues20}. Concerning illumination, the angles corresponding to the slits being smaller, one parallel light source is in this case sufficient to perform the measurement. The parameters of this configuration are perfectly suited for the 3D-tracking of microorganisms.

\section{Applications}

\subsection{Real-time label-free force spectroscopic measurement of the hybridization of short oligonucleotides with nanometric resolution.}

\begin{figure*}[!ht]
\centering
\includegraphics[width=18cm]{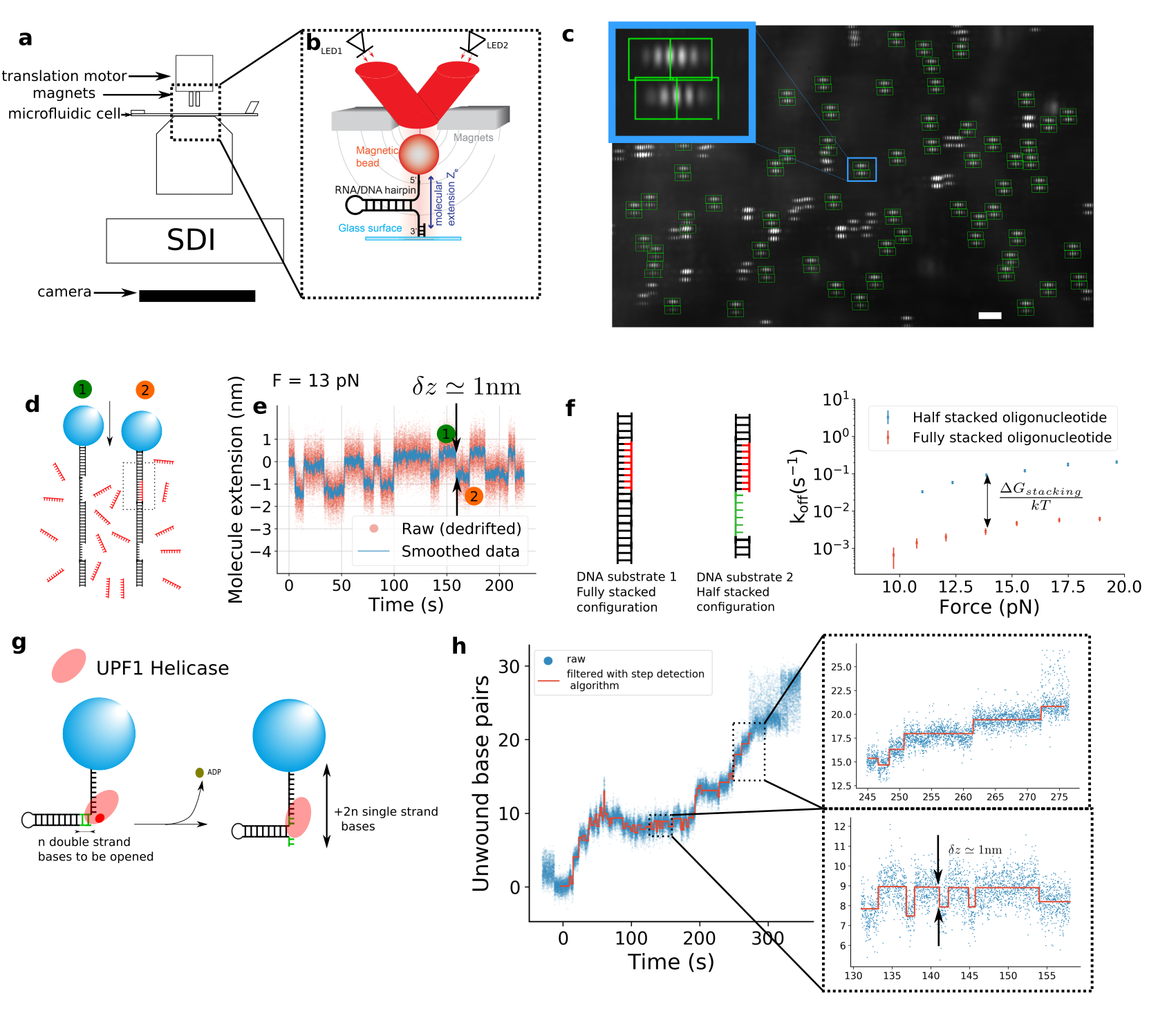} 
\caption{\textbf{a.} Schematic representation of the SDI setup attached to magnetic tweezers. \textbf{b.} In this picture, the bead is tracked with \emph{SDI}. Its \emph{z}-position is directly related to the number of bases hybridized in the hairpin. The magnets are designed so that both light sources go through the gap between them. \textbf{c.} A typical field of view. Each pair of fringes, materialized by two green boxes, corresponds to one magnetic bead. Scale bar, 10 $\mu$m. \textbf{d.} The hybridization of a 8-bp oligonucleotide causes a shortening of the DNA molecule of typically 1 nm since dsDNA is shorter than ssDNA above 5 pN. \textbf{e.} The 1-nm steps caused by the oligonucleotide hybridization are measured by tracking the position of the magnetic microsphere with the SDI. \textbf{f.} Two DNA substrates are tested with the same oligonucleotide : in the configuration on the left, there is no free base when the oligonucleotide hybridizes. The analysis of the kinetic parameters of the two-level systems described above allows accessing the stacking free energy by comparing $k_{off}$ and $k_{on}$ in both configurations. $k_{on}$ being equal for both configurations (see Figure~\ref{oligokon}), $k_{off}$ contains all the information about the free energy of stacking. \textbf{g.} Description of the helicase stepping experiment. As the helicase Upf1 unwinds a base pair of the hairpin, the measured extension increases by twice the length of a ssDNA base, that is roughly 0.9 nm at 9 pN. h) While unwinding the dsDNA recursively, Upf1 displays discrete steps (top inset). At the stalling position, Upf1 displays a ratchet-like behavior, going forth and back by steps of 1 bp (bottom inset).}
\label{Dessin2}
\end{figure*}

We now illustrate the precision of our tracking technique through various single-molecule force spectroscopy measurements. These measurements are performed with magnetic tweezers : DNA molecules are attached between a glass surface and a micrometric magnetic bead (Figure~\ref{Dessin2}.a-b). A force is applied by approaching magnets while the extension of the DNA molecule is measured by tracking the position of the bead with the help of \emph{Stereoscopic Darkfield Interferometry} (Figure~\ref{Dessin2}.c). The following examples are to our knowledge the first nanometric measurements of relevant biomolecular effects performed with a camera. They open the possibility to increase importantly the throughput of precise enzymatic measurements and to overcome the lack of parallelization that characterizes high resolution optical tweezers and AFM.

Figure~\ref{Dessin2}e shows the kinetics of hybridization of short oligonucleotides that were obtained by using the nanometric precision of \emph{SDI}. A DNA molecule with a short single stranded (ss) segment is attached between a magnetic microsphere (MyOne™) and the surface and pulled at a force of $\sim 13$ pN. A free 8-bp oligonucleotide complementary to the ssDNA segment of the tethered molecule is injected in the solution. As the oligonucleotide hybridizes to the molecule, 8 bases of the latter are converted from ssDNA to dsDNA, causing a shortening of the molecule of $\approx 1.5$ \AA \ per base at 13 pN, that is a total 1.2 nm extension difference (Figure~\ref{Dessin2}.d). When the oligonucleotide unbinds, the molecule recovers its original length. The resulting two-states (Figure~\ref{Dessin2}e) dynamics allow characterizing the kinetics of the hybridization process at difference forces between 10 pN and 20 pN (Figure~\ref{Dessin2}.f). 

 Base stacking is an interaction between the aromatic rings of successive nucleic acids and is a major contributor, alongside the hydrogen bonds between complementary bases, to the stability of the duplex structure \cite{yakovchuk2006base}.  We used the precision of \emph{SDI} to investigate the effect of the stacking of the oligonucleotide with the neighbour dsDNA. For this purpose, we reproduced the experiment on two different substrates (Figure~\ref{Dessin2}f). On one of them, the fully stacked configuration, the oligonucleotide hybridizes with a single stranded gap whose size is exactly 8 bp. On the other, the half-stacked configuration, the single-stranded gap is larger (14 bp) than the oligonucleotide, thus preventing stacking interaction on one of the free ends of the ss fragment. While a similar experiment was pioneered by Whitley \emph{et al.} in \cite{whitley_elasticity_2017-1} using FRET, we here are able to perform it in a fluorescence-free setup, thus avoiding an eventual bias that could be introduced by the interaction between the fluorescent label and the DNA. From the difference in the measured hybridization kinetics between the two substrates (Figure~\ref{Dessin2}f.), we notably extract the free energy of stacking (see Figure~\ref{oligostacked}) and find $\Delta G_{stacking,AG} = 2.0 \pm 0.2 \text{ kcal/mol}$. The free energy does not depend on the force in the considered range (12-18 pN). This is coherent with the fact that the elongation of dsDNA is almost constant in this force range. The method could be easily expanded to a more complete characterization of the energetics of such  short segments of nucleic acids.
 
 These nanometric experiments have a temporal resolution close to the second, the latter being limited by physical reasons related to the bead explained in \ref{noisebrenner}. For signals of larger spatial amplitude, the temporal resolution is drastically improved. For hairpin fluctuating between its open and close states, involving differences of extension between the two states of 10 nm, data show well-defined millisecond steps. (Figure~\ref{HP10}). 
 
\subsection{Sequential helicase stepping over a range of 30 base pairs.}

Helicases are essential enzymes that unwind dsDNA to separate the two DNA strands. They are key elements of DNA replication and repair. As an helicase unwinds a DNA hairpin pulled at 9 pN and tethered between a surface and a magnetic bead, the measured elongation increases by twice the length of a ssDNA base for each base pair unwound, that is roughly 1 nm (see Figure~\ref{Dessin2}g.). Distinguishing individual steps of an helicase with magnetic tweezers thus requires tracking the bead with subnanometric resolution.

Upf1 is an helicase belonging to the Super-Family 1 that we studied in previous works using magnetic tweezers \cite{fiorini2015human}\cite{kanaan2018upf1}. On Figure~\ref{Dessin2}h, we show that the resolution enhancement provided by \emph{SDI} allows resolving its individual steps, provided that the ATP concentration is low enough (500 nM). The 30-bp hairpin is open step-by-step until the remaining unwound part becomes too short and starts to oscillate spontaneously between its open and closed states under the influence of the applied force and of the thermal fluctuations (starting from 272s). At some positions, the enzyme displays a ratchet-like behavior with back-and-forth steps of one base pair. Away from this stalling position, the distribution of step sizes displays intermediate sizes between 1 bp and 2 bp (c.f. Figure~\ref{Step size distribution}). Such intermediate step sizes were already observed in optical tweezers experiments \cite{cheng2011single} and interpreted as an asynchronous release of both stands. Stalling positions of the enzyme are reproducible from beads to beads and can be easily identified even at higher ATP concentration. We show for example in Figure~\ref{Upf1blockpic} a trace of the helicase domain of yeast-Upf1 unwinding a 30-bp hairpin at a concentration of 10 uM ATP and the corresponding distribution blockage positions over 80 single-helicase events (Figure~\ref{Upf1blockhist}). Blockages are determined with the resolution of the base pair.

\subsection{3D-micrometric-tracking of dilute single cells.}

Usually, 3D tracking techniques of microorganisms \cite{taute_high-throughput_2015} are dependent on calibration libraries consisting of images taken at different vertical position. Interpolating these images allows retrieving the axial position of an object. The linearity of \emph{SDI} avoids the use of such lookup tables and thus greatly reduces image processing needed for 3D-tracking, while dark field illumination facilitates automatic detection. 

\begin{figure*}[!ht]
\centering
\includegraphics[width=18cm]{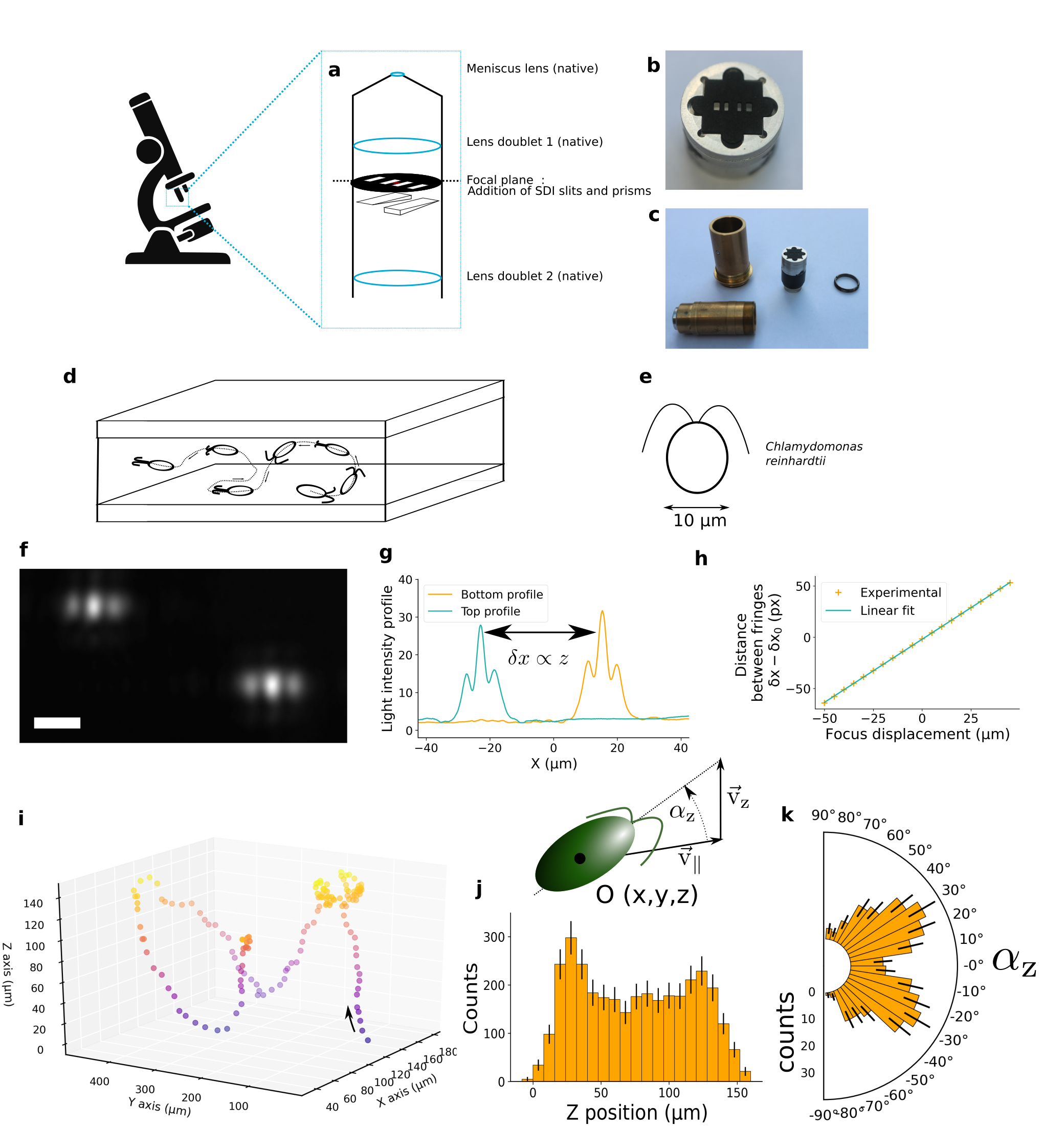} 
\caption{Direct integration of a modified objective in a microscope (here Olympus IX81) in order to implement \emph{Stereo Darkfield Interferometry} in an existing setup. \textbf{a.} Schematic description of the modified objective (Olympus, achromatic 20X). The \emph{SDI} slits brought in contact with two prisms of opposite angles are added in the Fourier plane of the objective. \textbf{b.} Picture of the slits and of the mechanics allowing their insertion in the objective. \textbf{c.} Picture of the disassembled objective.  \textbf{d.} Schematic representation of the tracking experiment : algae \emph {Chlamydonomonas reinhardtii} are inserted in a flow cell containing TAP buffer and their movement is tracked. \textbf{f.} An image of an \emph{alga} obtained with the integrated objective. White bar : 10 $\mu$m \textbf{g.} Horizontal profile of light intensity corresponding to the image $f$.The lateral shift between the two interference profile allows measuring the axial position $z$.  \textbf{h.} Dependency of the distance between the interference fringes on the position of the focus for an alga fixed on the surface. \textbf{i.} A 3D trajectory obtained thanks to the \emph{SDI} modified objective. Acquisition frequency : 10 Hz. Each point corresponds to one frame. Colors represent $z$. \textbf{j.} Distribution of the $Z$ position of the algae over 3314 positions taken from 117 individual trajectories.  \textbf{k.} Distribution of the vertical angles while an alga crosses the middle of the flow cell ($z$ between 60 and 90$\mu$m). 332 crossing events. 117 trajectories.}
\label{Dessin3}
\end{figure*}

Using a modified objective as described in the section \emph{Setup} enables the 3D-tracking of free micro-particles in any commercial microscope. We show in Figure~\ref{Dessin3}.f the image of the model micro-swimmer \emph{Chlamydomonas reinhardtii} acquired with the modified objective. 
We studied the dynamics of freely swimming algae confined in a microfluidic Hele-Shaw cell (i.e. width and length $\gg$ height) of thickness $\approx 150 {\rm \mu m}$. Filtering out short trajectories ( <1.5s spent in the field of view) and cells with deficient vertical swimming (spanning less than 110 um in the axial direction), we analysed the statistics of 117 independent trajectories and found that the dynamic is largely dominated by the presence of confining walls. As illustrated by the typical track in Figure~\ref{Dessin3}.i, most of the cells performed back and forth movements between the upper and lower boundaries (see also Figure~\ref{Vzvsz}). The distribution of their axial position displays peaks at $20-30 {\rm \mu m}$ from the cell limit (Figure~\ref{Dessin3}.j), in accordance with previous 2D horizontal measurements \cite{bruncosmebruny:tel-02501274}. Measuring the vertical angle $\alpha_z$ of each of the 117 trajectories within the middle of the chip (for $z$ between 60 and 90 ${\rm \mu m}$) we found that it is symmetrically distributed around 0 (Figure 3.k) with a clear peak at $\approx \pm \ang{25}$. This value compares reasonably well with the most probable outgoing angle following wall scattering as measured in \cite{kantsler2013ciliary}, which shows that after interacting with either wall the cells keep swimming in the same direction until colliding with the opposite boundary. This is consistent with the long $\approx 10$s run time extracted from the run-and-tumble dynamics performed by the algae in bulk \cite{Polin2009a}, since in our case it takes on average $\approx 2.5$s to cross the height of the channel. Finally, the symmetry observed in the distribution of angles indicates that gravity axis \cite{chlamigravitaxis} does not play any role in this confined configuration, simply because the gravitational torque \cite{Pedley1992} felt by the algae is too small to bias the cell swimming direction over such a short vertical distance.

%We tracked 332 trajectories of algae swimming in a microfluidic flow cell of height 150 $\mu$m.  After having dismissed short trajectories and algae that displayed particularly low swimming velocities (Figure~\ref{vdistrib}), we were left with 117 individual trajectories. The characteristics of the vertical movement of the algae was dominated by the confinement imposed by the height of the cell. We observed that the microorganisms performed back and forth movements between the upper and the lower limits of the cell (Figure~\ref{Dessin3}.i and Figure~\ref{Vzvsz}). The distribution of their axial position displays peaks at 20-30 $\mu$m from the chip boundaries, in accordance with previous 2D horizontal measurements \cite{bruncosmebruny:tel-02501274}.  We also measured the vertical angles of the 117 individual trajectories in the middle of the cell (for z laying between 60 and 90 $\mu$m). The latter are distributed symmetrically around 0 (Figure~\ref{Dessin3}.k), indicating that, contrarily to previous studies showing a strong bias towards the upper direction \cite{chlamigravitaxis}, gravitaxis does not dictate the orientation of the trajectories in such a confined configuration. On the other hand, the most probable angles (~$\pm25^{\circ}$) are larger than previously reported escape angles in 2D \cite{kantsler2013ciliary}, suggesting that surface interactions alone can not explain the observed vertical inclination of the algae. Further 3D-tracking experiments will be necessary to assess the effects of the coupled effect of microfluidic geometry, gravitaxis anf phototaxis.

\section{Discussion}

We presented a new 3D tracking method, \emph{Stereoscopic Darkfield Interferometry}. The darkfield allows for a reduction of the background noise, while the stereoscopic aspect allows for the linearity of the measurement and the interferometry ensures an Angström level resolution. By tracking the position of a magnetic bead with higher precision, the technique allows performing force spectroscopy experiments that were so far only possible with hardly parallelizable methods (AFM or optical tweezers) like helicase stepping measurements. It also enables the 3D-tracking of dilute free micro-organisms over a large range of axial positions without the need of any focus feedback nor any calibration. Because of the high information content of the PSF, the method can be performed at relatively low frequencies that are compatible with real-time tracking. 

Regarding its application to force spectroscopy, the method offers to the community working with  magnetic tweezers the means to study the discrete kinetic mechanisms of molecular motors. We showed some preliminary data on the helicase Upf1 but the technique could also be applied to the studies of other enzymes, like topoisomerases or polymerases. For example, the addition of one base by a polymerase results at 10 pN in the decrease of extension of the DNA molecule by a distance of roughly 1.5 \AA. Thus, our optical precision (a frame-by-frame standard deviation of 2 \AA \ in the axial direction due to photon noise) should theoretically allow the distinction of the discrete incorporation of bases by polymerases. However, there is still an obstacle to be overcome that is related to the Brownian motion of the magnetic beads. Indeed, while we measured Angström level standard deviation on the position of beads fixed on the glass surface, this noise increased to the nanometer range (at 160 Hz of acquisition frequency) when the beads were attached to a DNA substrate. This difference is partly due to the magnetic anisotropy \cite{mosconi2011soft} of the beads as pointed in \cite{klaue2009torsional} but mainly, it is due to the large increase of the axial hydrodynamic drag applied on the bead caused by the proximity of the surface \cite{brenner1961slow}. In our cases, the correction to the drag is close to factor 15 (see Figure~\ref{Brennercurve}), and thus implies an increase of the integrated Brownian motion of a factor $\sqrt{15} \simeq 4$. As we show in the Supplementary~\ref{noisebrenner}, this problem cannot be solved by using longer dsDNA handles as the induced loss of stiffness has a worse effect on the noise than the viscosity close to the surface. A technological development that could be transposed to magnetic tweezers in the future is the use of extremely stiff handles since they should allow a better spatial separation from the surface without having any notable impact on the stiffness of the construct.       

Regarding its applications to the tracking of micro-organisms, the method presents the advantage to be linear and thus to avoid the constitution of lookup tables.  However, it must be acknowledged that a large amount of light going through the sample (around 90\%) is absorbed by the black filter inside the SDI objective. It means that an intense light should be sent to the sample in order to get a bright enough signal. This can be problematic for the observation of light-sensitive species. In the case of \emph{Chlamydomonas}, the algae studied in this paper, this problem was overcome by using a wavelength (730 nm) that is thought to be large enough to not disturb the behavior of the organisms. 

The main point that needs to be carefully thought about in the process of developing new applications is the design of the slits, and the corresponding angle of the illumination. There are three parameters that have to be adjusted to the scatterer to be observed : their width, their length and their separation. For a given averaged angle, a larger separation between the slits allows for a better sensitivity and thus a better precision in the z-direction but decreases the depth of field. Larger widths and lengths (however fully contained inside the pupil of the objective) will increase the quantity of light that goes through the objective but will also reduce the width of the envelope, as well as reduce the depth-of-field of the signal. Once these three parameters are taken into account, the method is easy to learn and use, especially because it can be, as we showed, directly implemented to any existing microscope by engineering the objective. Furthermore, the signal analysis procedure, described in the Supplementary~\ref{algotrack} is simple and computationally inexpensive. For this reason, we claim that Stereo Darkfield Interferometry has the ability to extensively increase the adoption of 3D tracking methods in the biological community especially with the increase in quality and speed occurring in the camera field.

\section{Material and methods}
\subsection{Optics}
\subsubsection{Out-of-the-objective setup (40x and 100x objectives)}
Superluminescent LEDs (Exalos, EXS210030-03, 650 nm, 10 mW) are placed in the object focal plane of a diode collimation package (Thorlabs, LTN330A) and the outcoming parallel beams are directed to the pupil of the oil objectives (respectively UPLSAPO100X/UPLSAPO40X) with opposite incident angles. Their intensity is regulated through a feedback loop based on the signal of the monitor photodiode included in the chip. The optical axis is then switched from vertical to horizontal using a mirror (BBEA1-E0Z, Thorlabs). The 4f-setup , needed to have access to the Fourier planes as it is not possible with these objectives, is made of two lenses of focal 100mm (AC254, Thorlabs). A sheet of black paper (Canson, 160g/m2) is then cut with a laser machine in order to get the pair of slits needed for the SDI technique. Slits consist of four identical rectangles (200$\mu$mx700$\mu$m) whose centers are placed at the following positions with respect to the optical axis : ([-1.1mm,0],[-0.4mm,0],[0.4mm,0],[1.1mm,0]). The slits are brought in contact with a thin lens of focal 100mm (Thorlabs,  LBF254-100-A). Then, a pair of optically clear glass slides (4mmx8mm) of thickness 1mm is placed in contact with this lens. Both are tilted in the y-direction with opposite angles of $\pm 17$ degrees in order to shift the beams stemming from both pair of slits in opposite y-directions. Finally, the resulting pattern is imaged on a monochromatic CMOS camera (UI-3030CP-M, IDS Ueye), placed at 100mm from the titled glass slides. The camera is linked through USB3.1 to a computer that processes the images in real-time. Conjugation relations are shown in Figure~\ref{4f} and the protocol of the alignment is described in the Supplementary Materials~\ref{alignsi}.

\subsubsection{Modified 20x objective}

An achromatic objective 20X from Olympus was dismantled. We have replaced  a cylindrical spacer 14 mm wide and 10 mm long just after the front lens component of the objective by a holder supporting a black paper with slits followed by prisms and a spacer so that this system also spans 10 mm. A round glass slide of radius 13 mm was  inserted in the holder to hold the prisms. We have machined a (8 mm x 8 mm) rectangle slit in the holder allowing us to introduce two optically clear glass slides (4 mm x 8 mm) of thickness 1 mm in contact with the round one. One of the rectangular glass slides was slightly titled by introducing a thin sheet of paper of thickness $\approx 10 \mu m$ (cigarette paper ) between it and the round slide. The other rectangular glass slide was tilted in the opposite direction by introducing an identical sheet of paper on the opposite side. The space left between the round slide and the tiled rectangular slides was filled by capillarity with a drop of Olympus immersion oil, in order to create two prisms of opposite angles. A sheet of black paper in which the \emph{SDI} slits were cut were then added on top of the prisms.  
Slits consist in four rectangles which centers are placed at the following positions with respect to the optical axis : ([-2.2mm,0],[-0.9mm,0],[0.9mm,0],[2.2mm,0]). Outer slits (width : 0.85 mm, height : 1.25 mm) are wider than inner slits (width : 0.55 mm, height : 1.25 mm) in order to compensate for the smaller light intensity at large scattering angles.
The height of the cylindrical spacer was adjusted so that, once the objective is reassembled, the slits are located at the back focal plane of the objective.

\subsubsection{Measurement of the optical noise of the setup}
In order to measure accurately the optical noise of the setup and compare it to theory, it is necessary to uncouple it from the Brownian motion and from slow thermal or mechanical drifts that can affect the measurement. In order to do so, we melt MyOne T1 beads (ThermoFisher) on a glass surface. The stock solution was diluted 10000 times and then spread on a glass slide. The slide was then heated at 110$ ^{\circ}$C for 5 minutes following the evaporation of the solution. A standard flow cell was then assembled with this slide and water was injected. The position of 15 beads was then tracked for 5 seconds at 160 Hz at different light intensity. The average of the trajectory of the 15 beads was then subtracted from each individual trajectory in order to get rid of mechanical or thermal drifts.

\subsection{Temperature control}
In order to minimize optical drifts that are due to temperature changes, the objective is inserted into a box made of duralumin. The box temperature  is controlled through 4 Peltier moduli placed in parallel (ET-063-08-15, Adaptive). The PID feedback loop then allows to reach a temperature stabilization with the precision of 0.0001K. The temperature is read in the following way : thermistors (TDK, B57703M0103) are connected to Wheastone bridges  converting the resistance difference into a voltage. The voltage is then read with a 32-bit Sigma-Delta analog to digital converter (ADC).

\subsection{Focus adjustment}
For the linearity test of the setup with 40x or 100x magnification, the objective is moved with a piezo nanofocusing device (P-725.2CL, Physik Instrumente).
For the other measurements on this setup, the piezo nanofocusing device is replaced by a deformable stage to avoid any noise inherent to the piezoelectric device. The position of the screw controlling the stage is measured with a rotational magnetic sensor (AMS,AS5048A).
For the linearity test of the 20x setup used for algae tracking, the objective is moved by the stepper motor driving the microscope focus (IX81, Olympus).

\subsection{Single molecule assays}
All single-molecule experiments are performed at $25^{\circ}$C. All the DNA substrate used in the presented assays are synthesized single stranded oligonucleotides. Their sequences are indicated in Table~\ref{dnaseq}. Their 5'end is complementary to a 57 bases 3' DBCO modified long oligonucleotide (Oli1) that is attached to azide-functionalized surfaces (PolyAn 2D Azide) through a 2-hour-long incubation (100nM Oli1, 500 mM Nacl). Their 3'end is complementary to a 58-base oligonucleotide (Oli2). Oli2 contains two biotin modifications at its 5'end. The ssDNA substrate is first hybridized with Oli2 by mixing both oligos at 100nM in 100 mM Nacl, 30 mM, Tris-Hcl pH 7.6. 5uL of streptatividin coated Dynabeads MyOne T1 (Thermofisher) are washed three times in 200 uL of passivation buffer (140 mM NaCl, 3 mM KCl, 10 mM Na2HPO4, 1.76 mM KH2PO4, BSA 2\%, Pluronic F-127 2\%, 5mM EDTA, 10 mM NaN3, pH 7.4). The result of the hybridization between the substrate and Oli2 is diluted down to 2 pM buffer and then incubated 10 minutes with the beads in a total volume of 20uL of passivation buffer. The beads are then rinsed three times with passivation buffer in order to remove unbound DNA. 1uL of the bead solution is then introduced in the cell coated with Oli1 and filled with passivation buffer. They are incubated 5 minutes. Excess unbound beads are washed out by flowing passivation buffer.
\subsubsection{Hybridization assay}
The substrate used in these assays (HP10STACK) is a 153-base ssDNA strand. Once hybridized to Oli1 and Oli2, the size of the remaining single-stranded DNA reduces to 38 bp. 24 of these bases fold into a 10 bp hairpin with a 4-base apex loop. This hairpin is used to test that only one DNA molecule is bound to the beads by testing the fluctuation between the open and closed state at 10 pN. Once this is checked, the hairpin is blocked by a 16bp oligonucleotide (OliBlock-half-stack) for the half-stack configuration or by a 22bp oligonucleotide (Oliblock-full-stack) for the fully stacked configuration.  The remaining bases of single-stranded DNA contain the 8-base sequence that is complementary to the oligonucleotide whose hybridization kinetics is measured. The assay is realized in  Hybridization Buffer (100mM Nacl, 40mM Tris-Hcl pH 7.6).
\subsubsection{Upf1 stepping assays}
A 193-base ssDNA is used in this assay (HP30). Once hybridized to Oli1 and Oli2, the size of the single-stranded DNA reduces to 78 bases. Among them, a 30-bp is included with a loop of 4 bases. It is flanked by two handles of 7 successive thymines. Once the beads are attached to the surface of the cell, the buffer is changed to Upf1 buffer (100mM KCl, 3mM MgCl2, 40 mM Tris-Hcl pH 7.6). The closing and opening of the hairpin are tested using a force scan going from 20 to 8 pN. The force is then fixed to 8 pN on average and the helicase domain of yeast-Upf1 purified as described in \cite{kanaan2018upf1} is injected with ATP in Upf1 buffer  at a concentration of 10 nM unless otherwise mentioned.
\subsubsection{DNA hairpin folding assay}
The substrate used in this assay (HP10FLUC) is a 160-base long ssDNA strand. Once hybridized to Oli1 and Oli2, the size of the remaining single-stranded DNA reduces to 45 bases. 24 of these bases fold into a 10 bp hairpin with a 4-base apex loop. The kinetic of the folding and unfolding transitions of the hairpin at different forces are acquired in passivation buffer.

\subsubsection{Cultures of micro-organisms and microfluidic experiments}
Cultures of {\it Chlamydmonas reinhardtii} strain CC125 were grown axenically in a Tris-Acetate-Phosphate (TAP) medium at \ang{20}C under periodic fluorescent illumination ($\approx 100 {\rm \mu E.m^2.s^{-1}}$, cycle 16h light/8h dark). Cells were harvested in the exponentially growing phase, then centrifuged at 800 rpm. for 10 min and the supernatants replaced with fresh TAP. A dilute suspension ($\approx 10^{-5}\%$) was then loaded into a simple microfluidic chip ($L\approx 1$cm, $W\approx 3$mm and $H=150 {\rm \mu m}$; made with standard soft photo-lithography techniques) where the algae dynamic was acquired at $10$ fps (camera uEye UI-3000SE) on an inverted IX81 Olympus microscope with the SDI objective described in the main text and using a $730$nm LED for illumination in order to prevent phototactic responses from the cells.

\subsubsection{Image analysis}

Images are transmitted in real-time to a computer and analyzed on the fly at the frequencies indicated in the main text. 
For high-magnification and high-precision applications (force spectroscopy), images are analyzed through Fourier decomposition using the algorithm described in the Supplementary~\ref{algotrack}. 
In the case of \emph{algae} tracking, the position of the fringes is simply measured through the computation of the barycenter of their intensity in the $x$ and $y$ direction and the resulting signal is processed through a Finite Impulse Response (FIR) filter of width 0.8 s (8 frames).

\section{Authors contribution}

M.R. designed, performed and analyzed the biological experiments, performed the signal treatment and the theoretical analysis, contributed to the setup and wrote the manuscript. T.V. built, calibrated and characterized the optical setup and contributed significantly to its design. G.R built the electronics. T.V. and G.R. built the mechanics. N.R. expressed and purified Upf1. B.D. contributed to the hairpin constructions. R.J. performed the culture of \emph{Chlamydomonas reinhardtii} and supervised the analysis of their trajectories. V.C. supervised the research, designed the optical setup, its mechanics and its electronics and contributed to its experimental implementation. J-F.A. contributed to the design of the experiments and helped supervise the project. All authors discussed the results and commented on the manuscript. 

\section{Acknowledgement}

This work has originated in collaboration with Thomas Lepic. We wish to acknowledge stimulating discussions with H. Le Hir, J. Ouellet, David Bensimon, Nicolas Desprat, Ding Fangyuan, Maria Manosas  and their help with hairpins and ds-DNA constructs.

\section{Funding}

This study was supported by the ANR CLEANMD grant (ANR-14-CE10-0014), ANR G4-CRASH (G4-crash - 19-CE11-0021-01) from the French Agence Nationale de la Recherche to V.C., , by the European Research Council grant Magreps [267 862] to V.C.  and by continuous financial support from the Centre National de Recherche Scientifique, the Ecole Normale Supérieure  France and the Labex IPGG.

\section{Competing interests}
The authors declare the following competing interests : 
J-F.A. and V.C. own shares of the company $\text{Depixus}^\text{\textregistered}$ that makes a commercial use of \emph{Stereo Darkfield Interferometry}. 
T.V. and G.R. are now employed by the same company.
V.C. and T.V. and J-F.A have filed the patent US9933609B2 for the technique described in this paper. Application has been granted on 2018-04-03.
\end{multicols}

\bibliographystyle{unsrt}
\bibliography{These-papierSDI}

\begin{thebibliography}{10}

\bibitem{berg1972chemotaxis}
Howard~C Berg and Douglas~A Brown.
\newblock Chemotaxis in escherichia coli analysed by three-dimensional
  tracking.
\newblock {\em Nature}, 239(5374):500--504, 1972.

\bibitem{hodeib2016single}
Samar Hodeib, Saurabh Raj, M~Manosas, Weiting Zhang, Debjani Bagchi, Bertrand
  Ducos, Jean-Fran{\c{c}}ois Allemand, David Bensimon, and Vincent Croquette.
\newblock Single molecule studies of helicases with magnetic tweezers.
\newblock {\em Methods}, 105:3--15, 2016.

\bibitem{brouwer2020multiplexed}
Thomas~B Brouwer, Nicolaas Hermans, and John van Noort.
\newblock Multiplexed nanometric 3d tracking of microbeads using an fft-phasor
  algorithm.
\newblock {\em Biophysical Journal}, 2020.

\bibitem{dulin_high_2015}
David Dulin, Tao~Ju Cui, Jelmer Cnossen, Margreet~W. Docter, Jan Lipfert, and
  Nynke~H. Dekker.
\newblock High {Spatiotemporal}-{Resolution} {Magnetic} {Tweezers}:
  {Calibration} and {Applications} for {DNA} {Dynamics}.
\newblock {\em Biophysical Journal}, 109(10):2113--2125, November 2015.

\bibitem{huhle_camera-based_2015}
Alexander Huhle, Daniel Klaue, Hergen Brutzer, Peter Daldrop, Sihwa Joo, Oliver
  Otto, Ulrich~F. Keyser, and Ralf Seidel.
\newblock Camera-based three-dimensional real-time particle tracking at {kHz}
  rates and Ångström accuracy.
\newblock {\em Nature Communications}, 6(1), December 2015.

\bibitem{ragan_3d_2006}
Timothy Ragan, Hayden Huang, Peter So, and Enrico Gratton.
\newblock 3d {Particle} {Tracking} on a {Two}-{Photon} {Microscope}.
\newblock {\em Journal of Fluorescence}, 16(3):325--336, May 2006.

\bibitem{yildiz_kinesin_2004}
A.~Yildiz.
\newblock Kinesin {Walks} {Hand}-{Over}-{Hand}.
\newblock {\em Science}, 303(5658):676--678, January 2004.

\bibitem{abbondanzieri_direct_2005}
E.A. Abbondanzieri, W.J. Greenleaf, J.W. Shaevitz, R.~Landick, and S.M. Block.
\newblock Direct observation of base-pair stepping by {RNA} polymerase.
\newblock {\em Nature}, 438(7067):460--465, 2005.

\bibitem{greenleaf_direct_2008}
W.J. Greenleaf, K.L. Frieda, D.A.N. Foster, M.T. Woodside, and S.M. Block.
\newblock Direct observation of hierarchical folding in single riboswitch
  aptamers.
\newblock {\em Science}, 319(5863):630--633, 2008.

\bibitem{von_diezmann_three-dimensional_2017}
Alex von Diezmann, Yoav Shechtman, and W.~E. Moerner.
\newblock Three-{Dimensional} {Localization} of {Single} {Molecules} for
  {Super}-{Resolution} {Imaging} and {Single}-{Particle} {Tracking}.
\newblock {\em Chemical reviews}, 117(11):7244--7275, June 2017.

\bibitem{kim_high-resolution_2009}
Kipom Kim and Omar~A. Saleh.
\newblock A high-resolution magnetic tweezer for single-molecule measurements.
\newblock {\em Nucleic Acids Research}, 37(20):e136--e136, November 2009.

\bibitem{hanne_opening_2007}
Jeungphill Hanne, Giovanni Zocchi, Nikolaos~K. Voulgarakis, Alan~R. Bishop, and
  Kim~Ø. Rasmussen.
\newblock Opening rates of {DNA} hairpins: {Experiment} and model.
\newblock {\em Physical Review E}, 76(1):011909, July 2007.

\bibitem{brutzer_scanning_2012}
Hergen Brutzer, Friedrich~W. Schwarz, and Ralf Seidel.
\newblock Scanning {Evanescent} {Fields} {Using} a pointlike {Light} {Source}
  and a {Nanomechanical} {DNA} {Gear}.
\newblock {\em Nano Letters}, 12(1):473--478, January 2012.

\bibitem{graves_dynamic_2015}
Evan~T Graves, Camille Duboc, Jun Fan, François Stransky, Mathieu
  Leroux-Coyau, and Terence~R Strick.
\newblock A dynamic {DNA}-repair complex observed by correlative
  single-molecule nanomanipulation and fluorescence.
\newblock {\em Nature Structural \& Molecular Biology}, 22(6):452--457, June
  2015.

\bibitem{taute_high-throughput_2015}
K.~M. Taute, S.~Gude, S.~J. Tans, and T.~S. Shimizu.
\newblock High-throughput 3d tracking of bacteria on a standard phase contrast
  microscope.
\newblock {\em Nature Communications}, 6(1):1--9, November 2015.

\bibitem{gosse_magnetic_2002}
Charlie Gosse and Vincent Croquette.
\newblock Magnetic {Tweezers}: {Micromanipulation} and {Force} {Measurement} at
  the {Molecular} {Level}.
\newblock {\em Biophysical Journal}, 82(6):3314--3329, June 2002.

\bibitem{baddeley_three-dimensional_2011}
David Baddeley, Mark~B. Cannell, and Christian Soeller.
\newblock Three-dimensional sub-100 nm super-resolution imaging of biological
  samples using a phase ramp in the objective pupil.
\newblock {\em Nano Research}, 4(6):589--598, June 2011.

\bibitem{shechtman_optimal_2014}
Yoav Shechtman, Steffen~J. Sahl, Adam~S. Backer, and W.~E. Moerner.
\newblock Optimal point spread function design for 3d imaging.
\newblock {\em Physical Review Letters}, 113(13):133902, September 2014.

\bibitem{kao_tracking_1994}
H.~P. Kao and A.~S. Verkman.
\newblock Tracking of single fluorescent particles in three dimensions: use of
  cylindrical optics to encode particle position.
\newblock {\em Biophysical Journal}, 67(3):1291--1300, September 1994.

\bibitem{huang_three-dimensional_2008}
Bo~Huang, Wenqin Wang, Mark Bates, and Xiaowei Zhuang.
\newblock Three-dimensional super-resolution imaging by stochastic optical
  reconstruction microscopy.
\newblock {\em Science (New York, N.Y.)}, 319(5864):810--813, February 2008.

\bibitem{pavani_three-dimensional_2009}
Sri Rama~Prasanna Pavani, Michael~A. Thompson, Julie~S. Biteen, Samuel~J. Lord,
  Na~Liu, Robert~J. Twieg, Rafael Piestun, and W.~E. Moerner.
\newblock Three-dimensional, single-molecule fluorescence imaging beyond the
  diffraction limit by using a double-helix point spread function.
\newblock {\em Proceedings of the National Academy of Sciences of the United
  States of America}, 106(9):2995--2999, March 2009.

\bibitem{jia_isotropic_2014}
Shu Jia, Joshua~C. Vaughan, and Xiaowei Zhuang.
\newblock Isotropic 3d {Super}-resolution {Imaging} with a {Self}-bending
  {Point} {Spread} {Function}.
\newblock {\em Nature Photonics}, 8:302--306, 2014.

\bibitem{taylor2019interferometric}
Richard~W Taylor, Reza~Gholami Mahmoodabadi, Verena Rauschenberger, Andreas
  Giessl, Alexandra Schambony, and Vahid Sandoghdar.
\newblock Interferometric scattering microscopy reveals microsecond nanoscopic
  protein motion on a live cell membrane.
\newblock {\em Nature Photonics}, 13(7):480--487, 2019.

\bibitem{sun_parallax:_2009}
Yujie Sun, Jennine~Dawicki McKenna, John~M. Murray, E.~Michael Ostap, and
  Yale~E. Goldman.
\newblock Parallax: high accuracy three-dimensional single molecule tracking
  using split images.
\newblock {\em Nano Letters}, 9(7):2676--2682, July 2009.

\bibitem{sun_single-molecule_2010}
Yujie Sun, Osamu Sato, Felix Ruhnow, Mark~E. Arsenault, Mitsuo Ikebe, and
  Yale~E. Goldman.
\newblock Single-molecule stepping and structural dynamics of myosin {X}.
\newblock {\em Nature Structural \& Molecular Biology}, 17(4):485--491, April
  2010.

\bibitem{sancataldo2017three}
Giuseppe Sancataldo, Lorenzo Scipioni, Tiziana Ravasenga, Luca Lanzan{\`o},
  Alberto Diaspro, Andrea Barberis, and Mart{\'\i} Duocastella.
\newblock Three-dimensional multiple-particle tracking with nanometric
  precision over tunable axial ranges.
\newblock {\em Optica}, 4(3):367--373, 2017.

\bibitem{brevet}
Vincent Croquette, Jean-Fran{\c{c}}ois Allemand, and Thibault Vielle.
\newblock Optical device for measuring the position of an object, April~3 2018.
\newblock US Patent 9,933,609.

\bibitem{yakovchuk2006base}
Peter Yakovchuk, Ekaterina Protozanova, and Maxim~D Frank-Kamenetskii.
\newblock Base-stacking and base-pairing contributions into thermal stability
  of the dna double helix.
\newblock {\em Nucleic acids research}, 34(2):564--574, 2006.

\bibitem{whitley_elasticity_2017-1}
Kevin~D. Whitley, Matthew~J. Comstock, and Yann~R. Chemla.
\newblock Elasticity of the transition state for oligonucleotide hybridization.
\newblock {\em Nucleic Acids Research}, 45(2):547--555, January 2017.

\bibitem{fiorini2015human}
Francesca Fiorini, Debjani Bagchi, Herv{\'e} Le~Hir, and Vincent Croquette.
\newblock Human upf1 is a highly processive rna helicase and translocase with
  rnp remodelling activities.
\newblock {\em Nature communications}, 6(1):1--10, 2015.

\bibitem{kanaan2018upf1}
Joanne Kanaan, Saurabh Raj, Laurence Decourty, Cosmin Saveanu, Vincent
  Croquette, and Herv{\'e} Le~Hir.
\newblock Upf1-like helicase grip on nucleic acids dictates processivity.
\newblock {\em Nature communications}, 9(1):1--9, 2018.

\bibitem{cheng2011single}
Wei Cheng, Srikesh~G Arunajadai, Jeffrey~R Moffitt, Ignacio Tinoco, and Carlos
  Bustamante.
\newblock Single--base pair unwinding and asynchronous rna release by the
  hepatitis c virus ns3 helicase.
\newblock {\em Science}, 333(6050):1746--1749, 2011.

\bibitem{liphardt2001reversible}
Jan Liphardt, Bibiana Onoa, Steven~B Smith, Ignacio Tinoco, and Carlos
  Bustamante.
\newblock Reversible unfolding of single rna molecules by mechanical force.
\newblock {\em Science}, 292(5517):733--737, 2001.

\bibitem{woodside2006nanomechanical}
Michael~T Woodside, William~M Behnke-Parks, Kevan Larizadeh, Kevin Travers,
  Daniel Herschlag, and Steven~M Block.
\newblock Nanomechanical measurements of the sequence-dependent folding
  landscapes of single nucleic acid hairpins.
\newblock {\em Proceedings of the National Academy of Sciences},
  103(16):6190--6195, 2006.

\bibitem{bruncosmebruny:tel-02501274}
Marvin Brun-Cosme-Bruny.
\newblock {\em {Swimming of active suspensions in a crowded environment}}.
\newblock Theses, {Universit{\'e} Grenoble Alpes}, October 2019.

\bibitem{kantsler2013ciliary}
Vasily Kantsler, J{\"o}rn Dunkel, Marco Polin, and Raymond~E Goldstein.
\newblock Ciliary contact interactions dominate surface scattering of swimming
  eukaryotes.
\newblock {\em Proceedings of the National Academy of Sciences},
  110(4):1187--1192, 2013.

\bibitem{Polin2009a}
Marco Polin, Idan Tuval, Knut Drescher, J.~P. Gollub, and Raymond~E. Goldstein.
\newblock {Chlamydomonas Swims with Two "Gears" ina eukaryotic version of
  run-and-tumble locomotion}.
\newblock {\em Science}, 325(July):487, 2009.

\bibitem{chlamigravitaxis}
Vivian Kam, Nick Moseyko, Jeff Nemson, and Lewis~J. Feldman.
\newblock Gravitaxis in chlamydomonas reinhardtii: Characterization using video
  microscopy and computer analysis.
\newblock {\em International Journal of Plant Sciences}, 160(6):1093--1098,
  1999.

\bibitem{Pedley1992}
T.~J. Pedley and J.~O. Kessler.
\newblock {Hydrodynamic Phenomena In Suspensions Of Swimming Microorganisms}.
\newblock {\em Annual Review of Fluid Mechanics}, 24(1):313--358, 1992.

\bibitem{mosconi2011soft}
Francesco Mosconi, Jean~Fran{\c{c}}ois Allemand, and Vincent Croquette.
\newblock Soft magnetic tweezers: a proof of principle.
\newblock {\em Review of scientific instruments}, 82(3):034302, 2011.

\bibitem{klaue2009torsional}
Daniel Klaue and Ralf Seidel.
\newblock Torsional stiffness of single superparamagnetic microspheres in an
  external magnetic field.
\newblock {\em Physical review letters}, 102(2):028302, 2009.

\bibitem{brenner1961slow}
Howard Brenner.
\newblock The slow motion of a sphere through a viscous fluid towards a plane
  surface.
\newblock {\em Chemical engineering science}, 16(3-4):242--251, 1961.

\bibitem{bagchi2018single}
Debjani Bagchi, Maria Manosas, Weiting Zhang, Kelly~A Manthei, Samar Hodeib,
  Bertrand Ducos, James~L Keck, and Vincent Croquette.
\newblock Single molecule kinetics uncover roles for e. coli recq dna helicase
  domains and interaction with ssb.
\newblock {\em Nucleic acids research}, 46(16):8500--8515, 2018.

\bibitem{thompson_precise_2002}
Russell~E. Thompson, Daniel~R. Larson, and Watt~W. Webb.
\newblock Precise nanometer localization analysis for individual fluorescent
  probes.
\newblock {\em Biophysical Journal}, 82(5):2775--2783, May 2002.

\bibitem{opfer2012identifying}
Jan Opfer and Kay-Eberhard Gottschalk.
\newblock Identifying discrete states of a biological system using a novel step
  detection algorithm.
\newblock {\em PLoS One}, 7(11), 2012.

\end{thebibliography}

\newpage
\section{Supplementary materials}

\beginsupplement

\subsection{Conjugation relations in the optical setup of high magnification (100X and 40X)}
\begin{figure}[H]
\centering
\includegraphics[width=0.8\linewidth]{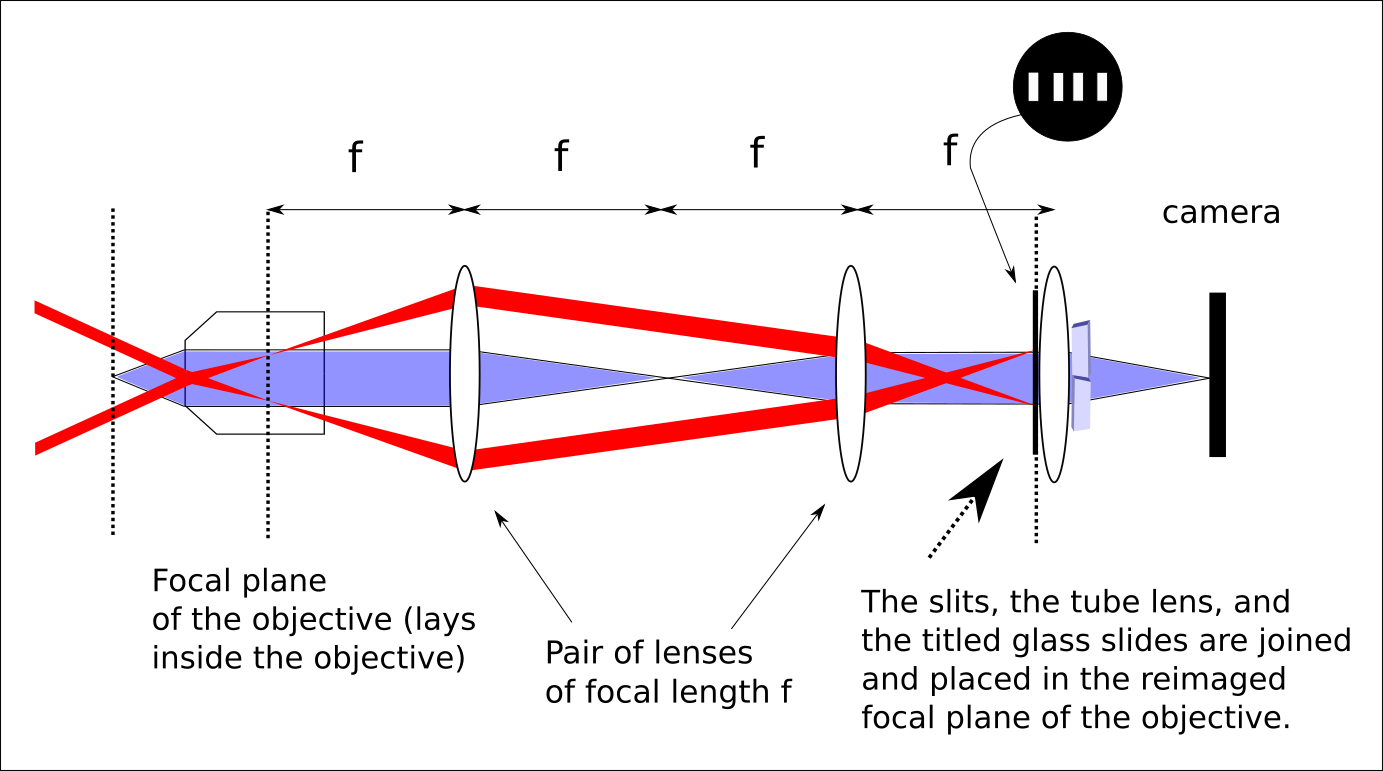} 
\caption{Scheme of the whole \emph{SDI} setup with all the optical elements being materialized. The pair of lenses after the objective is an afocal system that allows reimaging the Fourier plane of the objective.Red and blue dummy rays allow materializing conjugation relations.}
\label{4f}

\end{figure}

\begin{figure}[H]
\centering
    \includegraphics[width=0.8\linewidth]{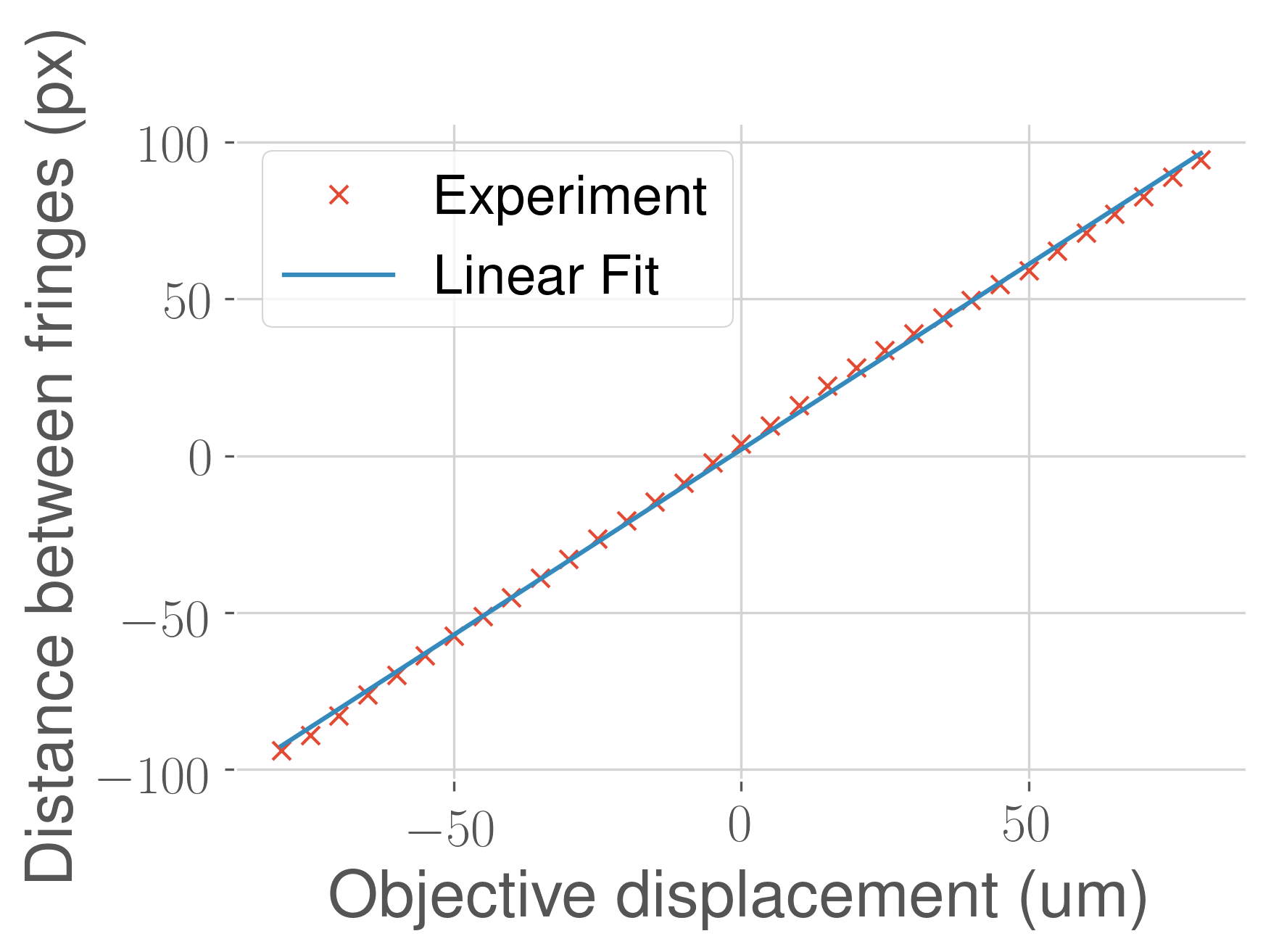}
    \caption{Dependence of the distance between interference fringes on the focal position of the emitter obtained by moving a 20X objective with a stepper motor.}
    \label{40_lin_whole}
\end{figure}

Usually, the object focal plane of an objective is not accessible as it is located inside the structure of the latter. In order to access the focal plane, a common way is to use a 4f-setup that allows reimaging it further away. Figure~\ref{4f} describes the whole setup. 

\subsection{SDI : depth of field and linearity}

\begin{figure}[H]
\centering
    \includegraphics[width=0.8\linewidth]{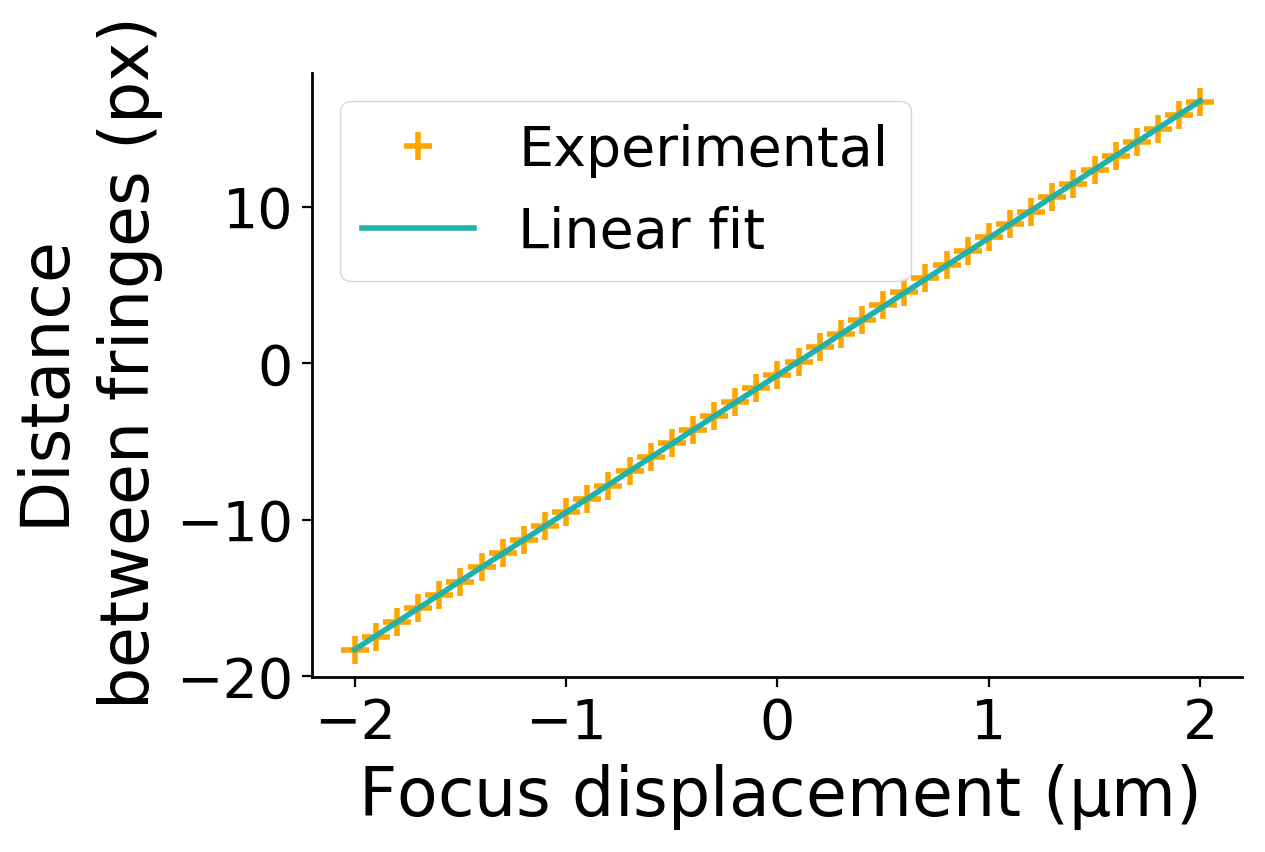}
    \caption{Dependence of the distance between interference fringes on the focal position of the emitter obtained by moving a 40X objective with a piezo stage.}
    \label{20_lin_whole}
\end{figure}

\begin{figure}[H]
\centering
    \includegraphics[width=0.8\linewidth]{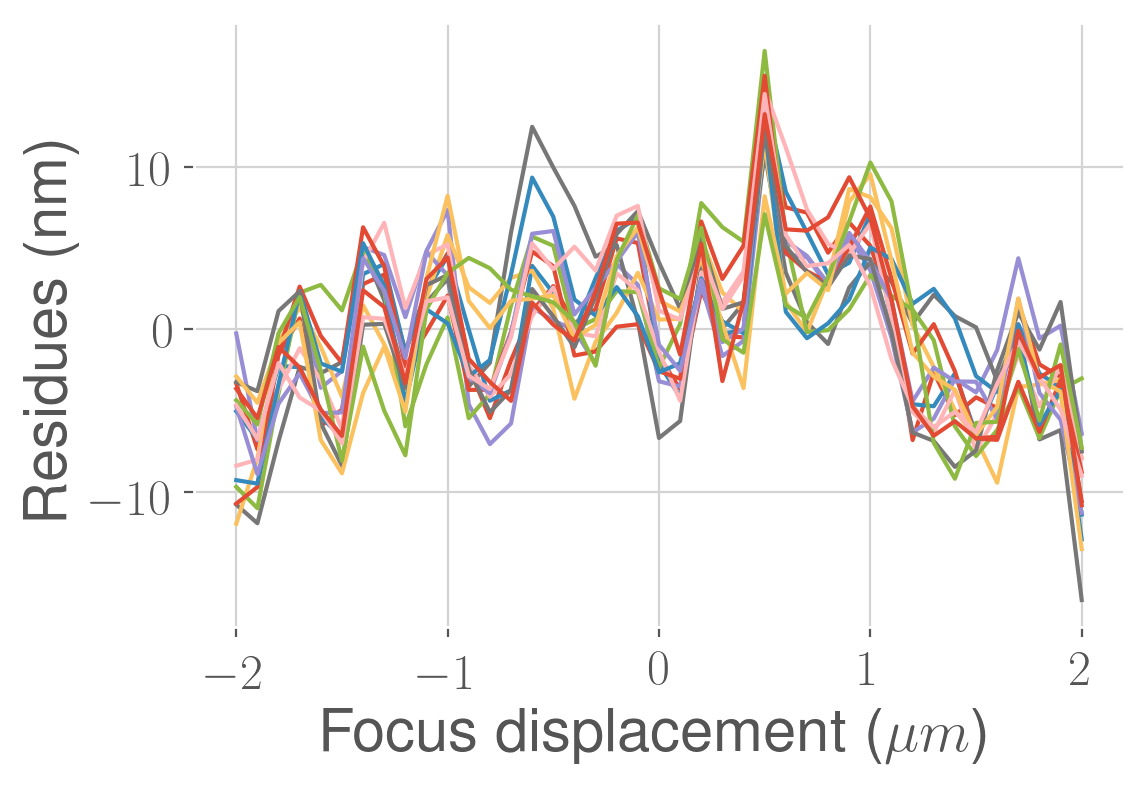}
    \caption{Residues from the linear fit of the dependence between the bead-to-objective axial (z) distance and the distance between the interference fringes for 16 different tracked beads. Objective 40X. }
    \label{residues}
\end{figure}

 \begin{figure}[H]
\centering
    \includegraphics[width=0.8\linewidth]{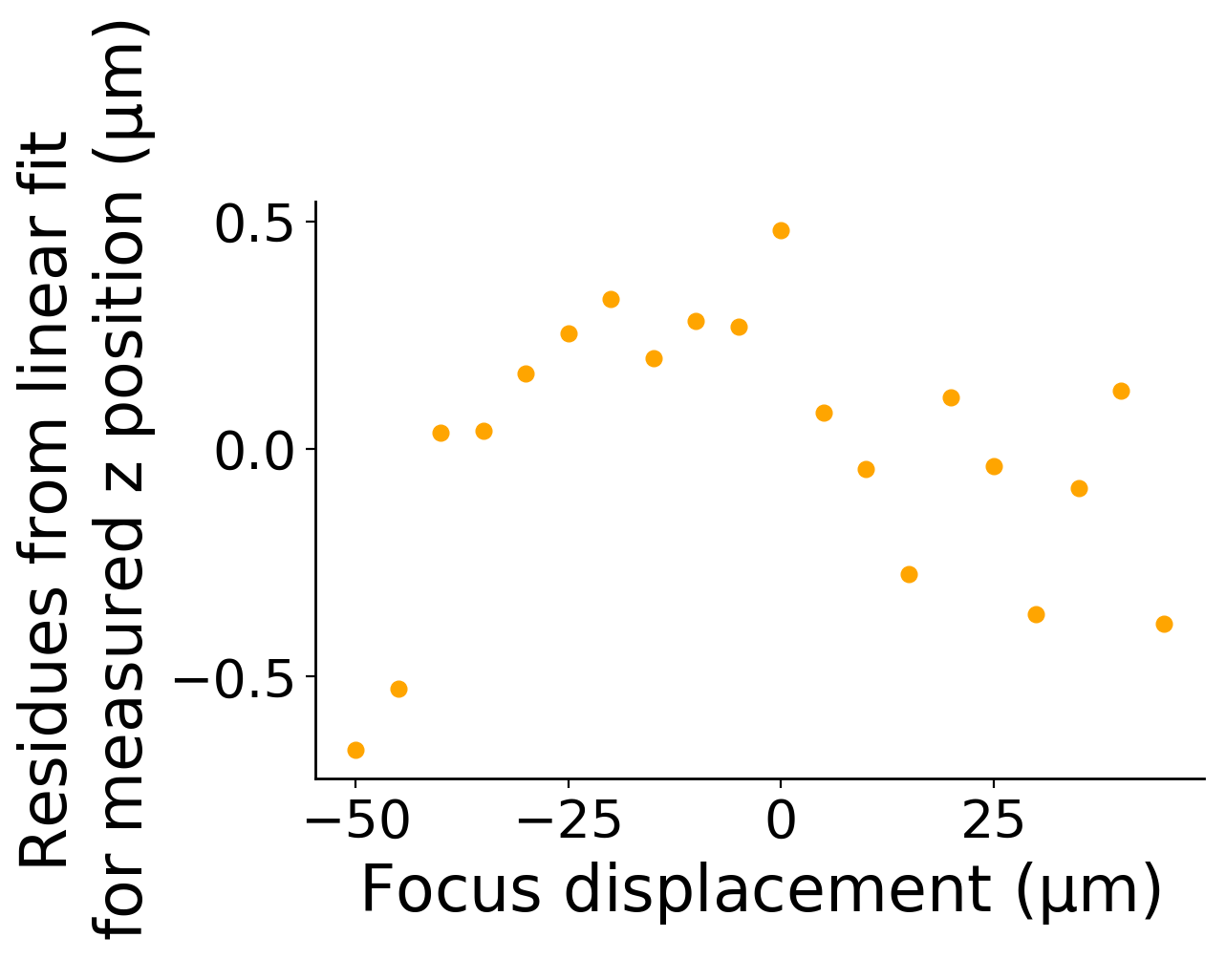}
    \caption{Residues from the linear fit of the dependence between the bead-to-objective axial (z) distance and the distance between the interference fringes. Objective 20X. }
    \label{residues20}
\end{figure}

\subsection{Theoretical precision}
\label{thprecision}
The precision of the localization of the particles depends on the minimal noise that can be obtained when measuring the translation of the interference patterns. Because the spread of the SDI Point Spread Function (PSF) is large compared to the pixel size $a$, we neglect in the following the error due to pixelization $\frac{a^2}{12}$ \cite{thompson_precise_2002}. 
The Cramér-Rao bound gives a theoretical lower bound for the standard deviation of the position measurement due to photon noise. The photon horizontal density profile depends on the translation parameter of the fringe $\epsilon$ in a simple way : 
\begin{equation}
f(x \mid \epsilon) = f(x - \epsilon)
\label{translationc}
\end{equation}
The Cramér-Rao bound stipulates that the  variance of the estimator of the hidden parameter $\epsilon$ from the sampling of the profile $f$ with $N_f$ photons verifies : 
\begin{equation}
    \sigma_{\epsilon_0}^2 > \frac{1}{N_f}\frac{1}{ - E \left( \frac{ \partial^2 \ln f(x\mid \epsilon)}{\partial \epsilon ^2}  \right) \Bigr|_{\substack{\epsilon_0}} }
    \label{cramer_rao_whole}
\end{equation},
where the expectation value E() is taken with respect to $f(x \mid \epsilon)$ and $\epsilon_0$ is the true value of the parameter to be estimated. The expectation value in the denominator is the Fisher information about the parameter $\epsilon$ contained in one photon.
Using equation~\ref{translationc}, it is straightforward to check that in the case of  the estimation of a translation parameter, $\sigma_{\epsilon_0}$ does not depend on the true value of the parameter $\epsilon_0$ and that equation~\ref{cramer_rao_whole} becomes 

\begin{equation}
\sigma_{x,pattern}^2 > \frac{1}{N_f}\frac{1}{ \bigintss \frac{f'(x)^2}{f(x)} dx}
\end{equation}

Equivalently, the precision of the localization of one fringe pattern in the $y$ direction is given by  :

\begin{equation}
\sigma_{y,pattern}^2 = \frac{1}{N_f}\frac{1}{ \bigintss \frac{f'(y)^2}{f(y)} dy} =  \frac{w_y^2}{N_f} 
\end{equation},
where the last equality is due to the fact that the y-profile is Gaussian of waist $w_y$.

We computed above the Cramér-Rao bound in the image space. The corresponding bounds in the object space can be deduced from the transverse magnification $g$ and the axial sensitivity $g_z$ that is related to $g$ through equation~\ref{axial_sensitivity}.

In order to get the transverse position of the bead, the positions of both interference pattern need to be averaged, and then divided by the magnification $g$:
\begin{equation}
\sigma_{x,bead}^2 = \frac{1}{2 g^2 N_f} \frac{1}{ \bigintss \frac{f'(x)^2}{f(x)} dx} = \frac{1}{g^2 N_{tot}} \frac{1}{ \bigintss \frac{f'(x)^2}{f(x)} dx} 
\end{equation}
\begin{equation}
\sigma_{y,bead}^2 = \frac{w_y^2}{2 g^2 N_f} = \frac{w_y^2}{ g^2 N_{tot}} 
\end{equation}
$N_{tot}$ being the total number of photons scattered (in both fringes pattern). 

The axial position of the bead is given by the difference between the position between the two fringes, and thus the minimal localization error in this direction reads :

\begin{equation}
\sigma_{z,bead}^2 = \frac{1}{2 g_z^2 N_f} \frac{1}{ \bigintss \frac{f'(x)^2}{f(x)} dx} = \frac{1}{g_z^2 N_{tot}} \frac{1}{ \bigintss \frac{f'(x)^2}{f(x)} dx} = \frac{\sigma_{x,bead}^2}{\tan^2(\frac{\alpha_1+\alpha_2}{2})}
\end{equation}

These theoretical errors are compared with experimental ones on the figure~\ref{Setup} in the main text. 

\begin{table}[!ht]
    \centering
\begin{tabular}{|l|c|c|c|}
  \hline
  Imaging profile & Fisher & Maximal  & Theoretical \\
   & Information & photon number & localization \\
   & (px\textsuperscript{-2}.photon\textsuperscript{-1}) & per fringe and per frame & precision (px) \\
   
  \hline
  SDI pattern & \cellcolor{green} $0,18$ & \cellcolor{green} $7.4  \times 10^6$ & $0.9 \times 10^{-3}$ \\
  One fringe & \cellcolor{green} $0,23$ & \cellcolor{orange} $1.8  \times 10^6$ & $1.6 \times 10^{-3}$ \\
  Envelope & \cellcolor{orange} $0,006$ & \cellcolor{green}  $11  \times 10^6$ & $3.9 \times 10^{-3}$  \\
    \hline
 
\end{tabular}
 \caption{Comparison of information related properties of the different profiles presented in Figure~\ref{Setup}.}
 \label{tablenoise}
 \end{table}
 
 \subsection{Computation of axial magnification}
\label{grandissement}
\begin{figure}[H]
\centering
\includegraphics[width=1\linewidth]{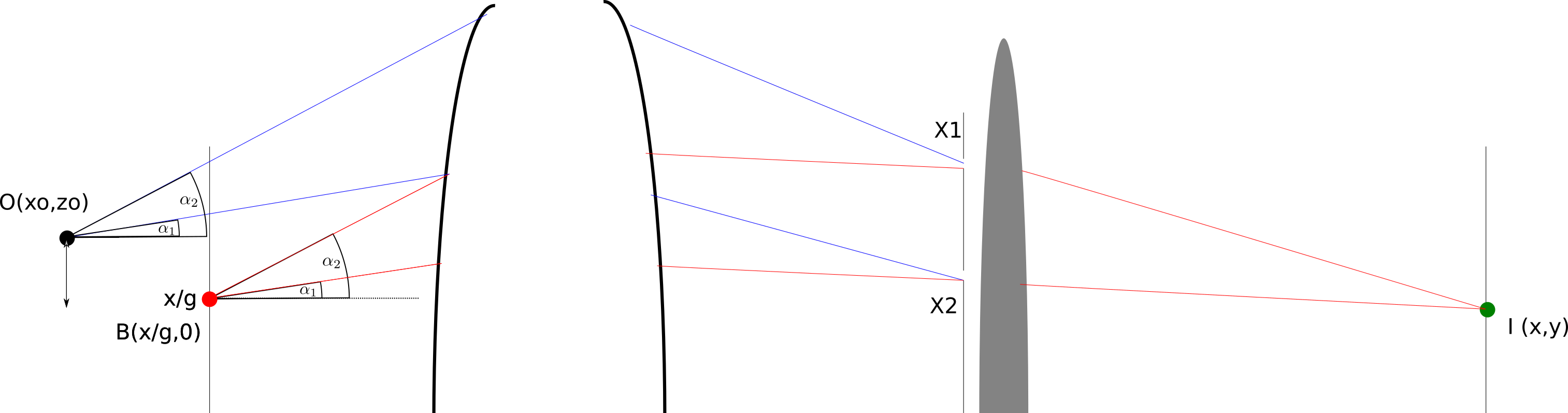} 

\end{figure} 

\begin{figure}[H]
\centering
\includegraphics[width=0.8\linewidth]{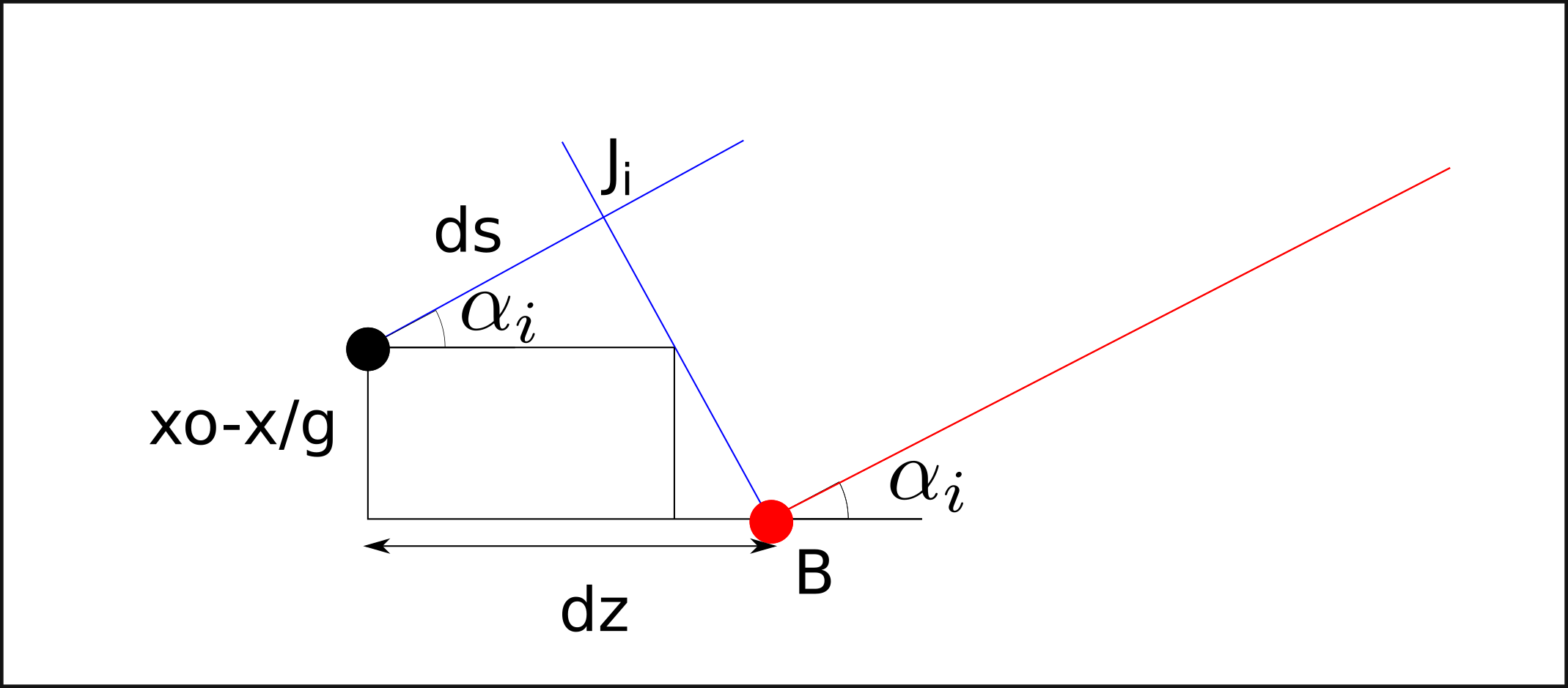} 
\caption{\textbf{Top :} Figure defining the notations used to calculate the transverse magnification of the setup. The object to be imaged is placed in O. The interferences are observed in I, the plane of the camera, \emph{i.e.} the back focal plane of the setup. I is conjugated to the point B, placed in the front focal plane of the setup. The $\alpha_i$ are the angles of the rays going through the slits at the position $X_i$. \textbf{Bottom : } Figure defining the construction point $J_i$, belonging to the same wavefront as B with respect to the wave of angle $\alpha_i$.}

\label{ds}
\end{figure} 

In this part we compute the phase shift between two rays stemming from the object $O(x_o,z_o)$ when they reach a point of the screen $I(x,z)$ by assuming that the objective is perfectly aplanetic but without assuming the astigmatism along the axial coordinate. For this purpose we use a construction point belonging to the object focal plane whose image would be I. Given the hypothesis of perfect aplanetism, the coordinates of this point are $B(x/g,0)$, where g is the transverse magnification of the setup and all the rays stemming from this point and converging to I have no relative phase shift. Thus, denoting $\alpha_1$ et $\alpha_2$ the angles corresponding to the slits, we have :
\begin{equation}
(BI)_{\alpha_1} = (BI)_{\alpha_2} 
\end{equation},
so,
\begin{equation}
(BX_1) + (X_1 I) = (BX_2) + (X_2 I)
\end{equation}
Incoming parallel rays being conjugated to the plane of the slits, the following relation also holds :
\begin{equation}
(J_1 X_1) = (BX_1)
\end{equation},
and,
\begin{equation}
(J_2 X_2) = (BX_2)
\end{equation},
$J_1$ being defined figure~\ref{ds} and $J_2$ being the corresponding construction point for $\alpha_2$. They verify (see figure~\ref{ds}):
\begin{equation}
(OJ_i) = \cos(\alpha_i)z_o - \sin(\alpha_i)(x_o - \frac{x}{g})
\end{equation}
Bringing these relations together, we get
\begin{equation}
\begin{split}
\delta(OI) & = (OI)_{\alpha_1} - (OI)_{\alpha_2} \\
& = (OJ_1) + (J_1X_1) + (X_1 I) - (OJ_2) - (J_2X_2) - (X_2 I) \\
& = (OJ_1) - (OJ_2) \\
& =   ( \cos(\alpha_1) -  \cos(\alpha_2)) z_o - (\sin(\alpha_1)-\sin(\alpha_2)) (x_o - \frac{x}{g}) 
\end{split}
\end{equation}
Thus,
\begin{equation}
\begin{split}
\delta \phi & = \frac{2\pi n \delta(OI)}{\lambda} \\
& = \frac{2\pi n}{\lambda} ( \cos(\alpha_1) -  \cos(\alpha_2)) z_o - (\sin(\alpha_1)-\sin(\alpha_2)) (x_o - \frac{x}{g}) 
\end{split}
\end{equation}
We deduce in particular the lateral displacement of the first fringe when the bead is displaced of $\delta z_b$ in the axial direction :
\begin{equation}
dx = g \frac{\cos(\alpha_1) -  \cos(\alpha_2)}{\sin(\alpha_2)-\sin(\alpha_1)} \delta z_b
\end{equation}
Thus, the axial magnification is written (recalling that $g$ is the transverse magnification) :
\begin{equation}
g_z = g \frac{\cos(\alpha_1) -  \cos(\alpha_2)}{\sin(\alpha_2)-\sin(\alpha_1)} = g \tan(\frac{\alpha_1+\alpha_2}{2})
\label{axial_sensitivity}
\end{equation} 
We also deduce the interfringe : 
\begin{equation}
i = g\frac{\lambda}{n(\sin(\alpha_2) - \sin(\alpha_1))}
\end{equation}

 \subsection{Localization algorithm}
\label{algotrack}

As seen in the previous section, the localization of the bead in the three directions is based on the precise sub-pixel localization of the interference fringes in the X and Y. Here we detail the algorithms of localization.

\subsubsection{X localization}

\begin{figure}[H]
\centering
    \includegraphics[width=0.8\linewidth]{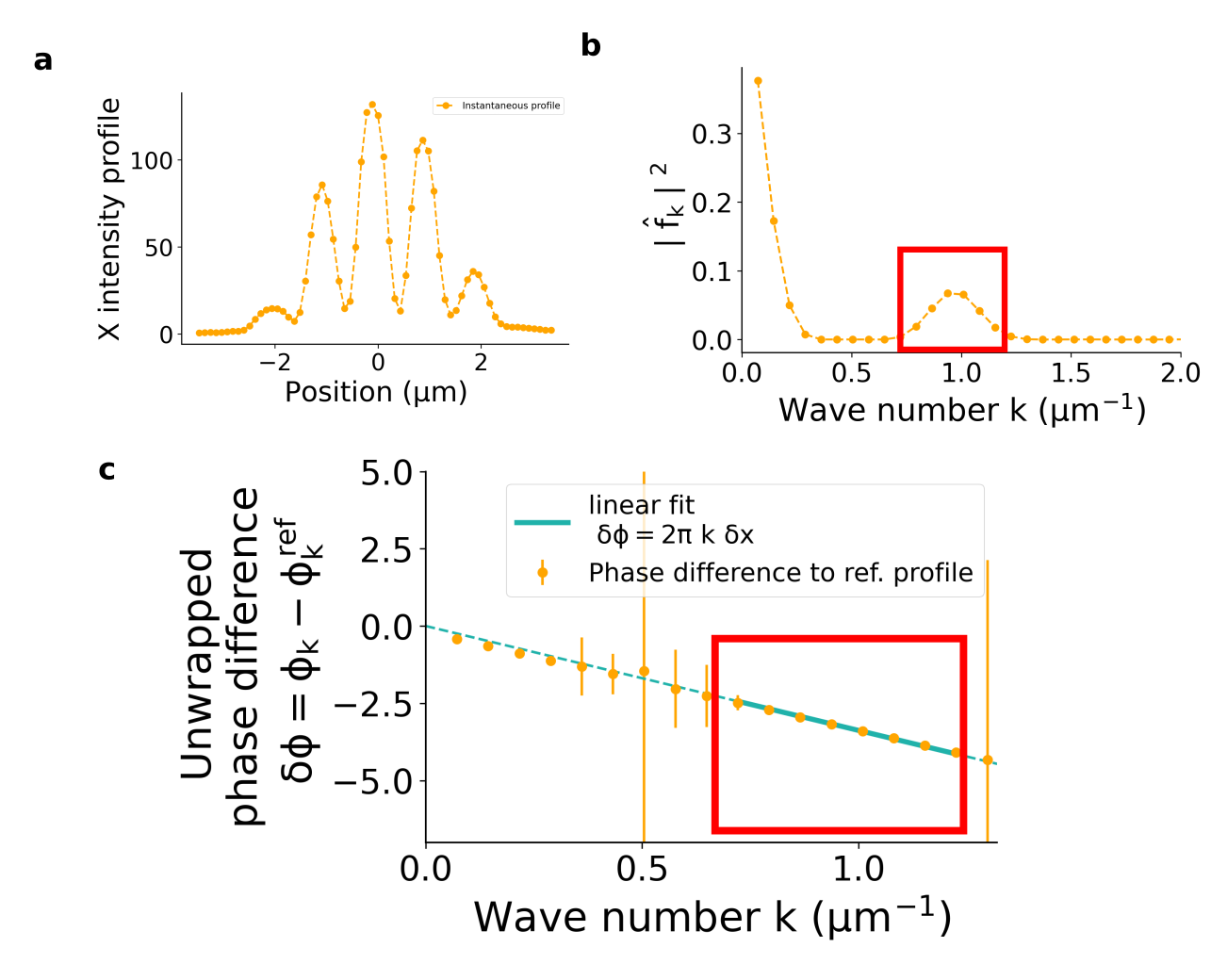}
    \caption{\textbf{a.} Instantaneous intensity X-profile of an interference pattern. \textbf{b.} Amplitude of the Fourier modes of the pattern showed in a. The sum is normalized to 1. The red square shows the modes that correspond to the oscillating part of the pattern. \textbf{c}. Unwrapped phase shift between the reference profile and the instantaneous profile showed in a. The phases of the modes inside the red square are fitted by a linear function. The displacement of the pattern in the direction $x$ is inferred from the slope.  Errors on the phase are taken as the inverse of the Fourier modulus shown in b.}
    \label{Fourierspectre}
\end{figure}

The translation of a pattern in X is measured through the following steps :
\begin{enumerate}
    \item A Blackmann-Nuttal window of width 128 pixels is applied on the pattern.
    \item The windowed signal $f(x)$ is decomposed on the Fourier basis using discrete Fourier transform (DFT). Figure~\ref{Fourierspectre}.b shows the amplitude $\mid \hat{f}_k \mid ^ 2$ of the Fourier modes. The envelope corresponds to the slow modes while the oscillating part of the signal is contained in the second peak of the spectrum.
    \item As the signal translates of a length $\delta x$, the phase of each mode $k$ is shifted by a quantity $2\pi k \delta x$. This property is used in order to infer $\delta x$. Using a fixed reference profile $f_x^{ref}$, we compute the phase shift of each mode between $f^{ref}$ and $f$ : $\delta \phi_k = \phi_k^{ref} - \phi_k$.
    \item $\delta \phi_k$ is eventually unwrapped and its values at the modes of interest are fitted by a linear function (see Figure~\ref{Fourierspectre}.c). The value of the x-displacement of the pattern $\delta x$ is inferred from the slope $a$ of the fitted linear function: $\delta x = \frac{a}{2\pi}$.
    
\end{enumerate}

\subsubsection{Y localization} 

\begin{figure}[H]
    \centering
    \includegraphics[width=0.8\linewidth]{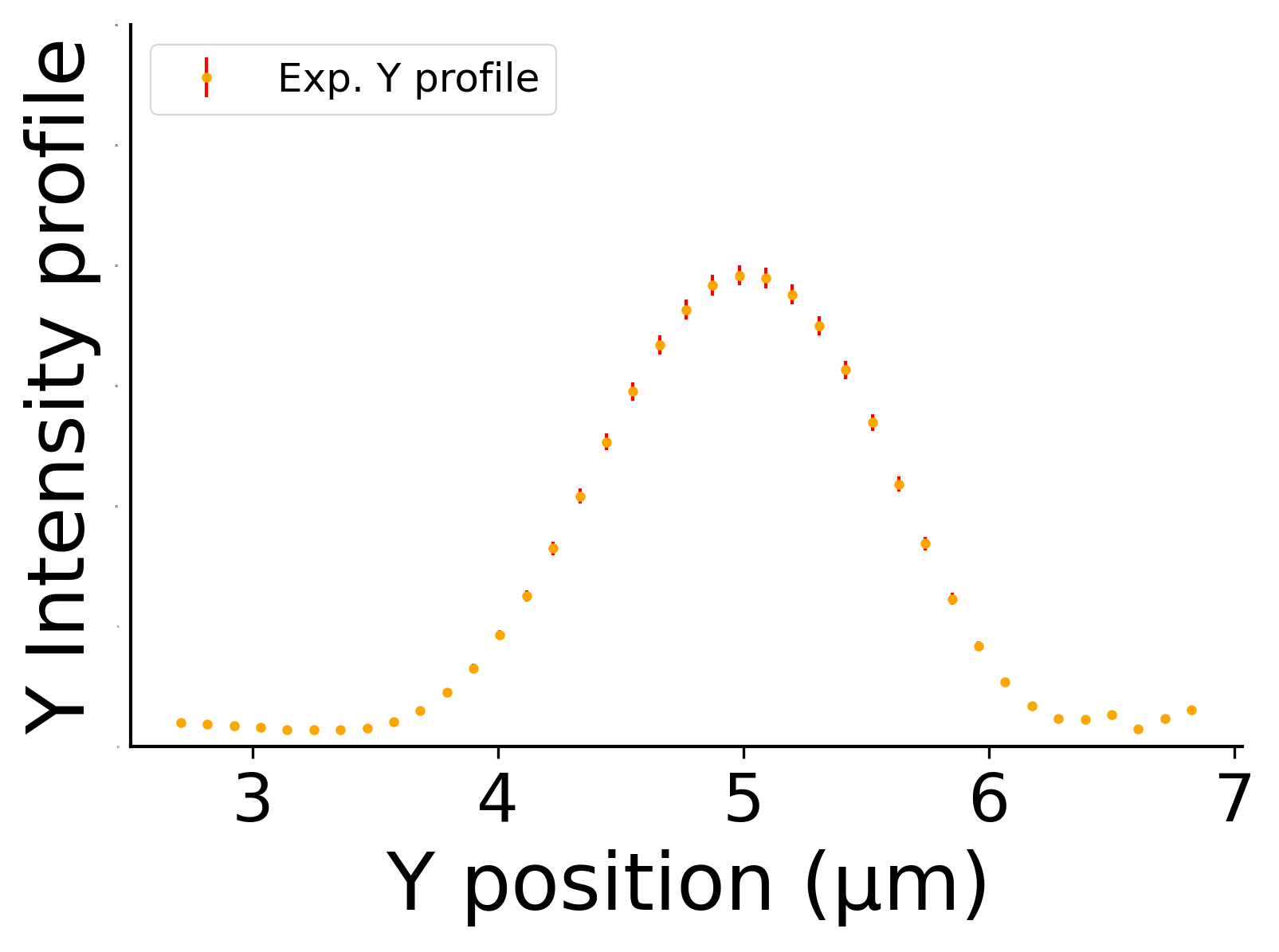}
    \caption{Intensity profile of the \emph{SDI} pattern in the Y direction.  Relative errors are equal to $\frac{1}{\sqrt{N}}$, where N is the number of photons received in each pixel. }
    \label{Yalgo}
\end{figure}

Figure~\ref{Yalgo} shows the intensity profile in the Y direction of \emph{SDI} pattern. In order to avoid the need for background correction, the center of the profile is obtained by computing the barycenter $\sum_{i=1}^n i f(x_i)$, where $i$ is stands for the pixel and f for the normalized light intensity.

\subsubsection{Comparison to theoretical precision}

\begin{figure}[H]
    \centering
    \includegraphics[width=0.8\linewidth]{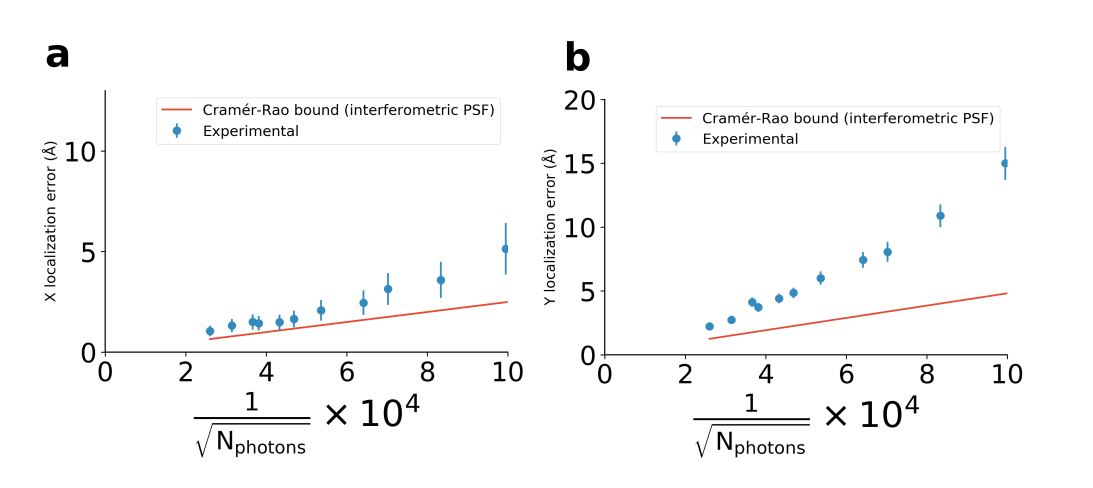}
    \caption{Experimental localization error in the X  (\textbf{a}) and the Y (\textbf{b}) directions are compared to the Cramér-Rao lower bound. Protocols for the measurements are the same as presented in the legend of  Figure~\ref{Setup} and in the section \emph{Material and Methods}. }
    \label{CRLBalongXY}
\end{figure}

The previously described algorithms are evaluated with regard to the theoretical precision (Cramér-Rao bound) expected for the \emph{SDI} pattern in Figure~\ref{CRLBalongXY}. The X-localization almost performs as well as expected by the theory.  While the Y-localization also allows reaching subnanometric localization error, there is a significant discrepancy with the theoretical prediction. This might be due to non-linear effects related to the oscillations of the Airy function that cannot be easily filtered out by Fourier decomposition, in opposition with the case of the X-direction where the fast modes can be isolated from slowly varying optical artefacts.

 \subsection{Oligonucleotide hybridization assay}
\begin{figure}[H]
\centering
\includegraphics[width=0.8\linewidth]{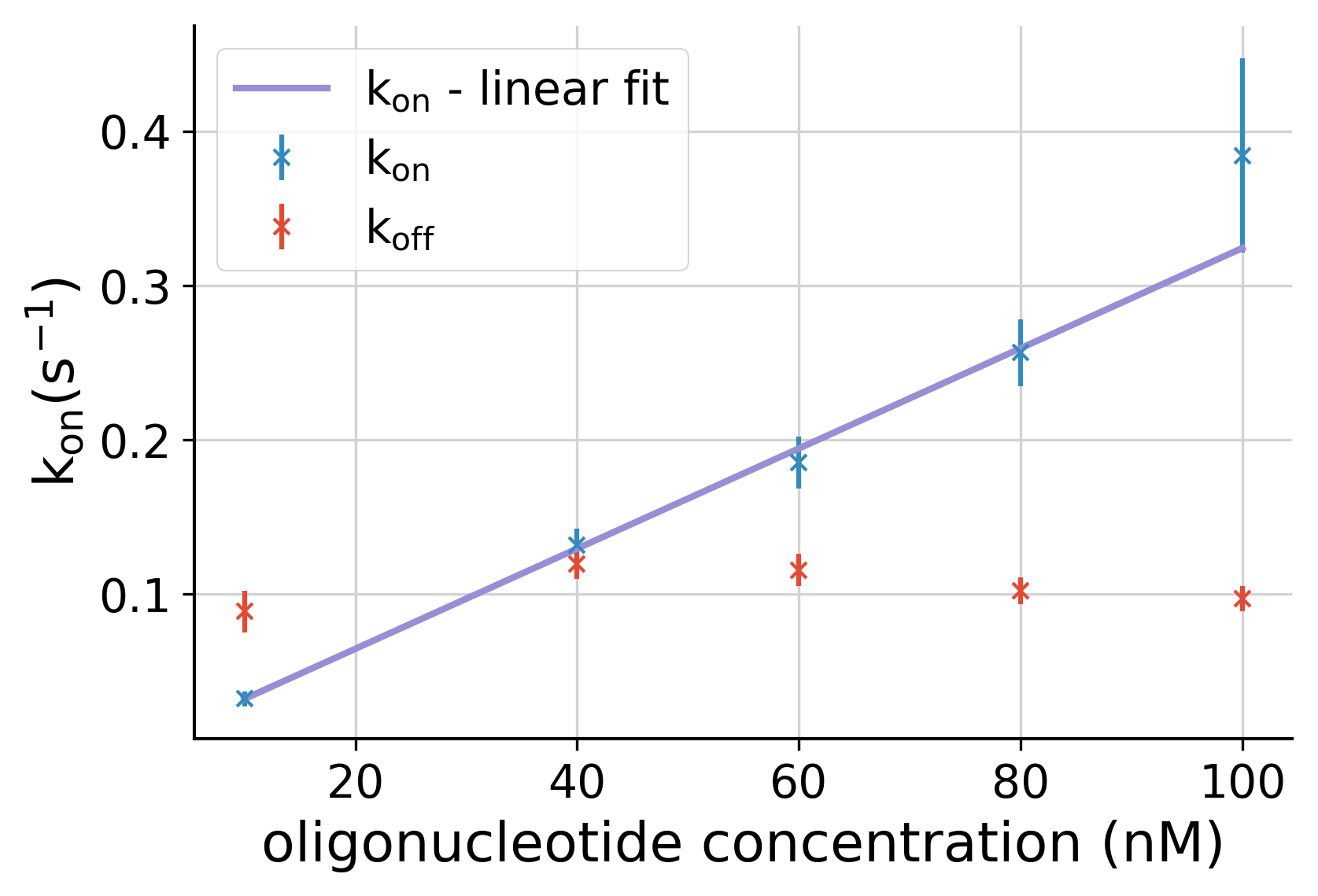} 
\caption{$k_{off}$ and $k_{on}$  of the oligonucleotides binding as a function of the concentration of oligonucleotide in the half-stacked configuration. The linearity of $k_{on}$ as a function of the concentration confirms the nature of the steps observed in the data.}
\label{oligoconcentrationtest}

\end{figure} 
\begin{figure}[H]
\centering
\includegraphics[width=0.8\linewidth]{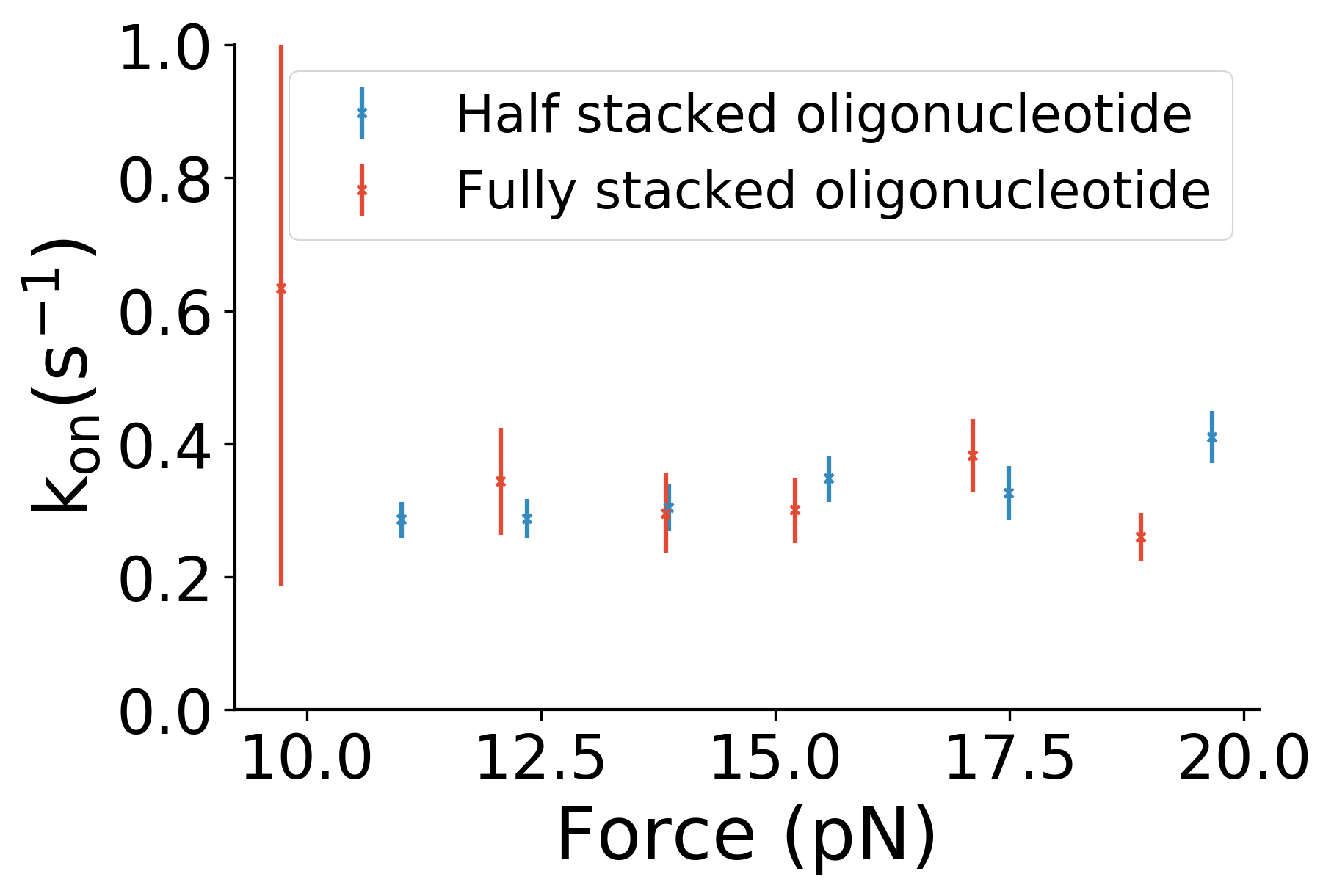} 
\caption{$k_{on}$  of the oligonucleotides binding as a function of applied force in both configurations, half-stacked and fully stacked. No significant dependence on the force can be observed, confirming previous results, like in \cite{whitley_elasticity_2017-1}. }
\label{oligokon}

\end{figure}

 \begin{figure}[H]
\centering
\includegraphics[width=0.8\linewidth]{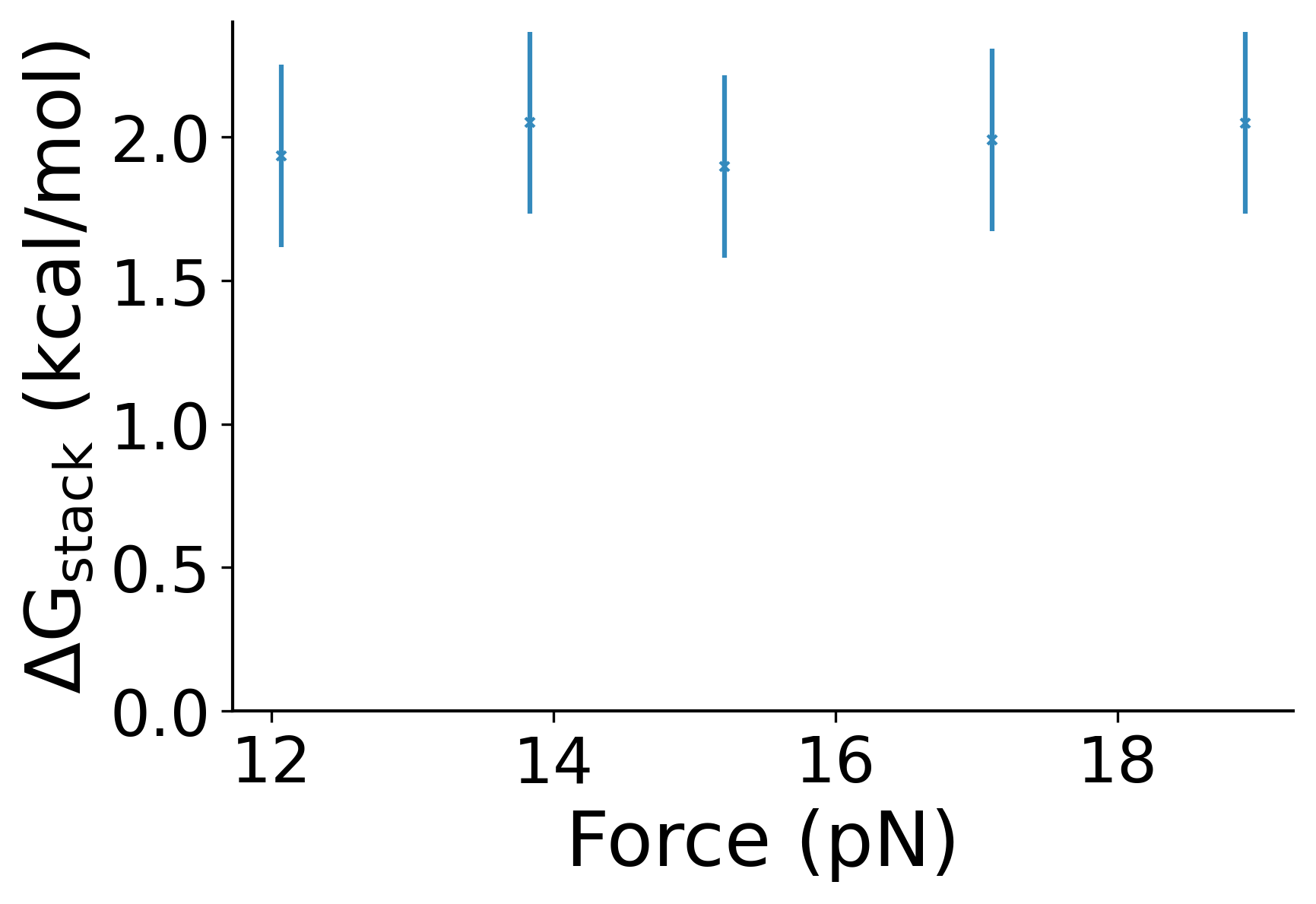} 
\caption{Dependence of the stacking free energy with the force. The free energy is computed in the following way : $\Delta G_{stacking} = k T \log \frac{k_{off,FS}k_{on,HS}}{k_{off,HS}k_{on,FS}}$, where FS and HS stand respectively for fully-stacked and half-stacked configuration. T = 298.15 K}
\label{oligostacked}

\end{figure} 

\subsection{Upf1 helicase}

 \begin{figure}[H]
\centering
\includegraphics[width=0.8\linewidth]{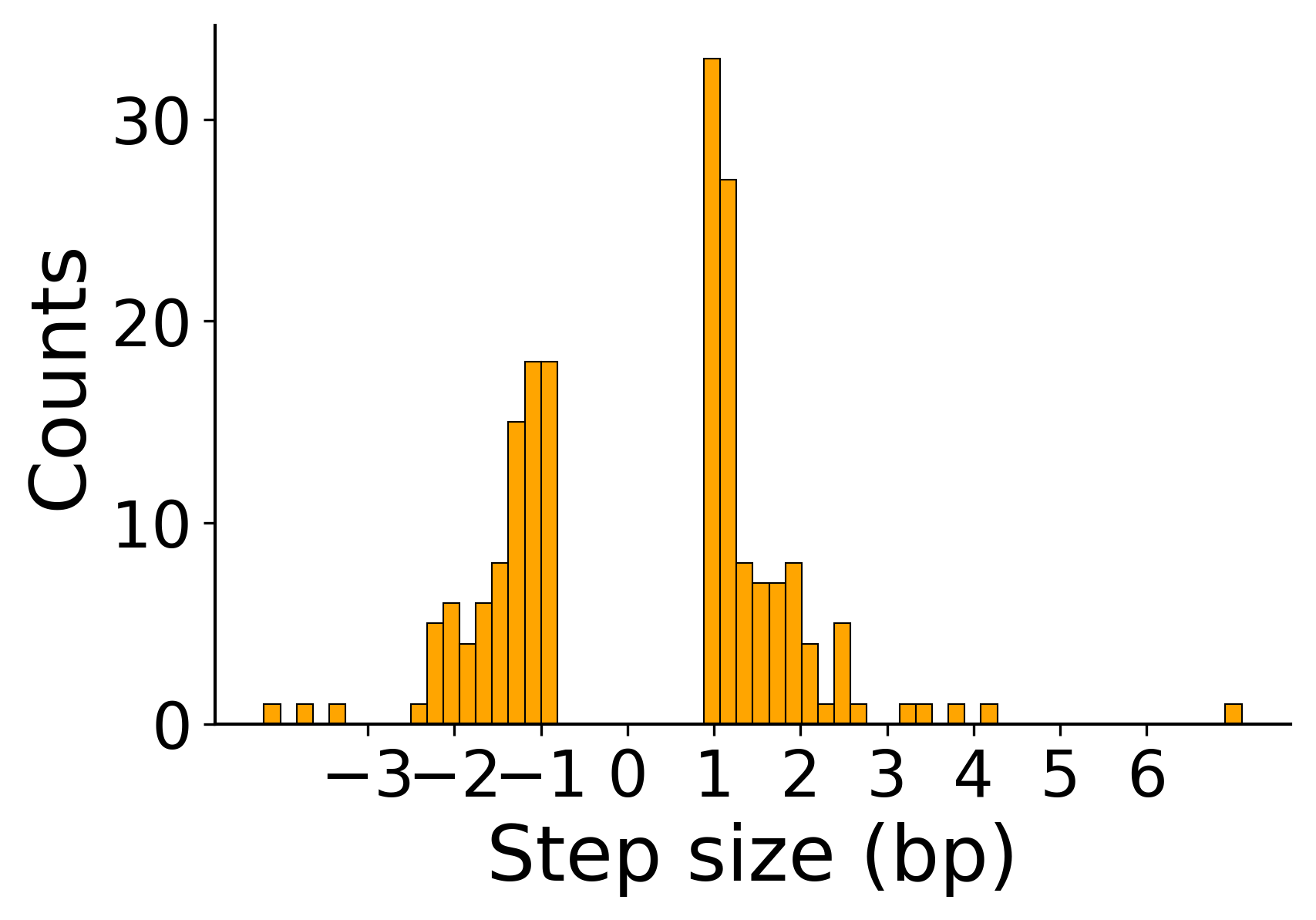} 

\caption{Distribution of measured step size over 190 stepping events during Upf1 unwiding at 500 nM ATP.}
\label{Step size distribution}

\end{figure}

\begin{figure}[H]
\centering
\includegraphics[width=0.8\linewidth]{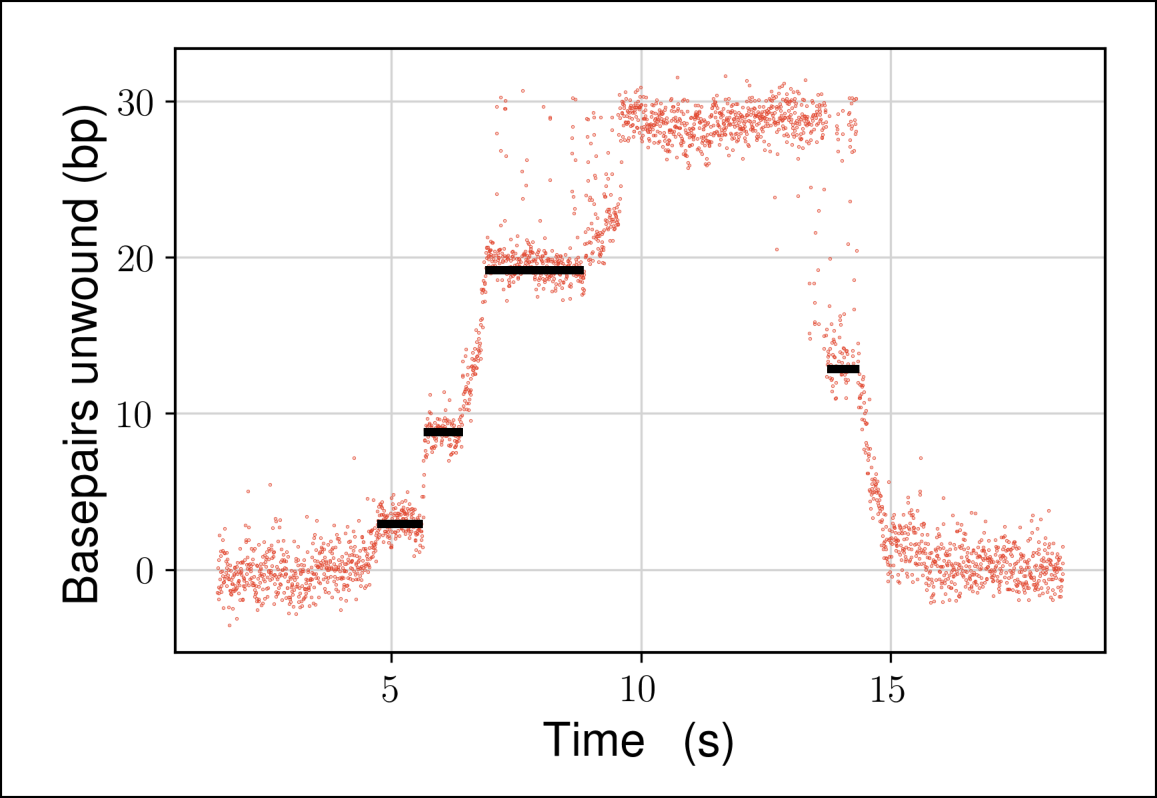} 
\caption{Trace showing the unwinding and the reclosing of a 30-bp Hairpin by the helicase domain of yeast-Upf1. The black lines show identified blockage. A blockage is identified with the following empirical criterium : The measured position must lie within a range of 2bp during at least 0.5s.  }
\label{Upf1blockpic}

\end{figure} 

\begin{figure}[H]
\centering
\includegraphics[width=0.8\linewidth]{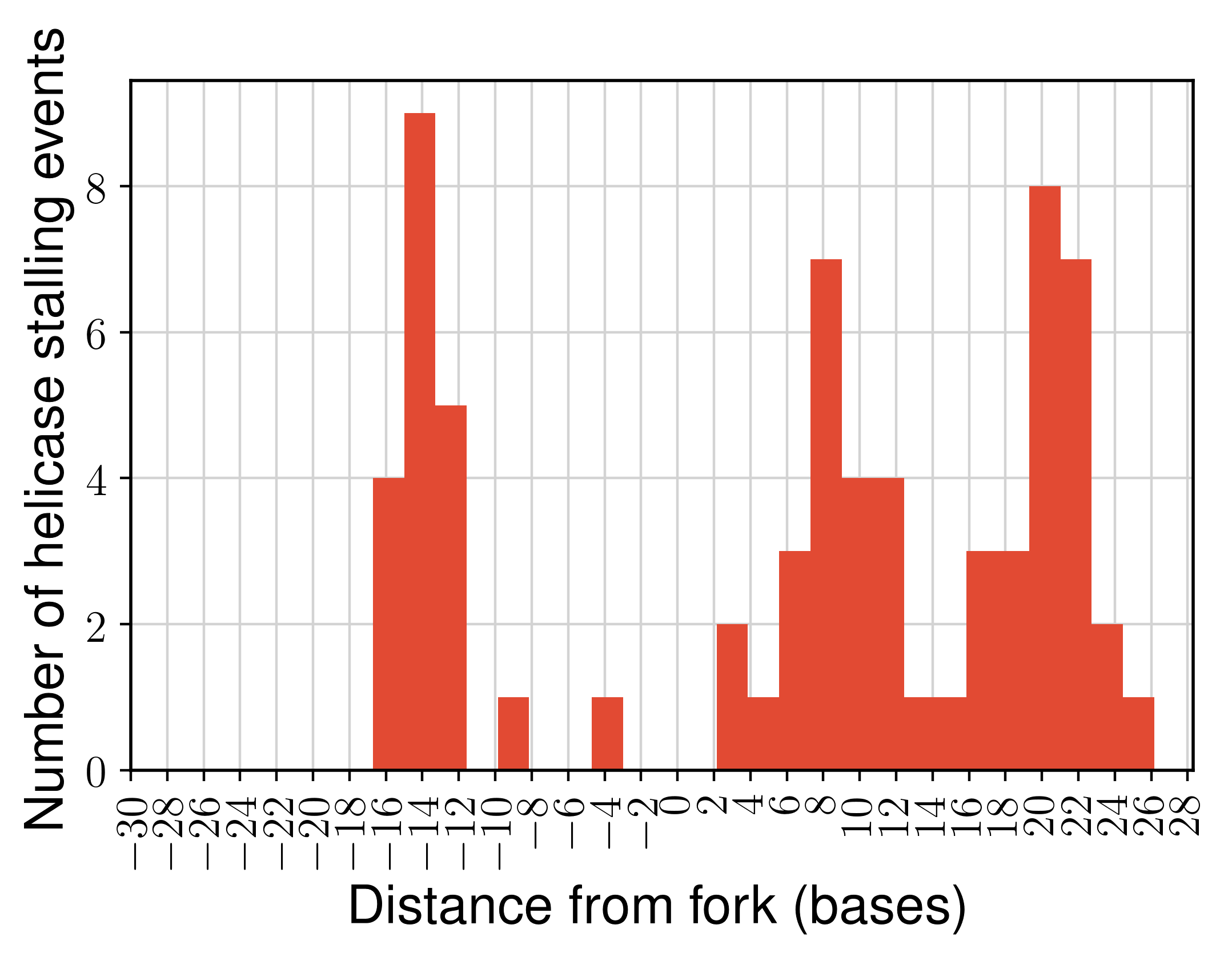} 

\caption{Distribution of the stalling positions of 80 stalling events of yeast-Upf1 identified as explained in the previous figure. Negative positions correspond to blockages happening during the closing of the hairpin while positive positions correspond to blockages happening during the opening of the hairpin.}
\label{Upf1blockhist}

\end{figure}  

\subsection{Step detection}

In order to detect steps in noisy data, we use the algorithm \emph{Moving Step Fit} (MSF) described in \cite{opfer2012identifying}, fitting the signal by horizontal lines instead of fitting it by any linear function.
A window $w$ is defined. For each point in the data $x_i$, the mean of the signal $m^l_i$ between $i-w/2$ and $i$ is computed along with the residues of the fit, called left residues $RSS_l$:
\begin{equation}
    RSS^l_i = \sum_{j=i-w/2}^i (x_j-m^l_i)^2
\end{equation}
Equivalently, the mean of the signal $m^r_i$ between $i$ and $i+w/2$ is computed along with the residues of the fit, called right residues $RSS_r$:
\begin{equation}
    RSS^r_i = \sum_{j=i}^{i+w/2} (x_j-m^r_i)^2
\end{equation}
Finally, the mean of the signal  $m_i$ over the whole window $w$ is computed, along with the whole residues $RSS$: 
\begin{equation}
    RSS_i = \sum_{j=i-w/2}^{i+w/2} (x_j-m_i)^2
\end{equation}
The MSF score is defined as :
\begin{equation}
    MSF_i =\mid m^r_i-m^l_i \mid (RSS_i -RSS^r_i - RSS^l_i)
\end{equation}
A step is detected at the position $x_i$ if the following conditions are true :
\begin{itemize}
    \item The step is larger than a given threshold $t_s$ : 
    \begin{equation}
        \mid m^r_i -m^l_i \mid > t_s
    \end{equation}
    \item The score $MFS_i$ is larger than a score threshold $t_{MFS}$.
\end{itemize}
We set the score threshold $t_{MFS}$ by computing the distribution of scores on a signal that precedes the injection of the helicase, \emph{i.e.} where no step is expected.  The threshold is defined as the score value where the cumulative probability of scores on such a step-free trace equals 99.99\%. For such a threshold, a false positive detection every 10000 points (62.5 seconds at 160 Hz) is expected. 

The parameters used to detect steps in the helicase data described in this paper are the following :
\begin{itemize}
    \item The window $w$ equals 100 points (0.8 seconds at an acquisition frequency of 160 Hz). 
    \item The step threshold $t_s$ equals 0.3 nm.
    \item The score threshold $t_{MFS}$ equals $1\times10^{-8} \mu m^3$
\end{itemize}

\begin{figure}[H]
\centering
\includegraphics[width=0.8\linewidth]{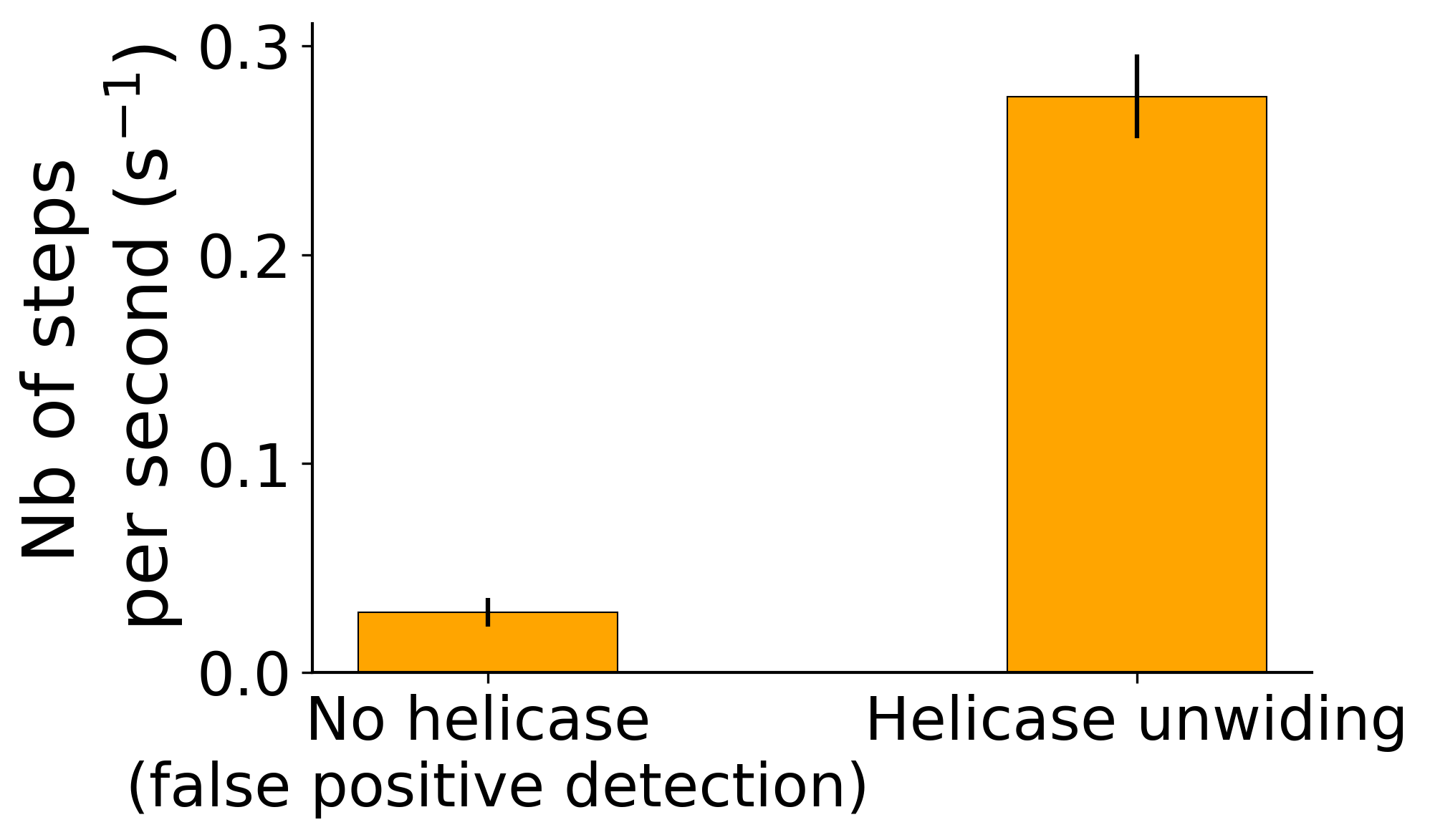} 

\caption{Left : Rate of false positive detection (steps detected without helicase). Nsteps=18, T = 628s. Right : Rate of step detection (steps detected during helicase unwiding). Nsteps = 190s, T = 689s }
\label{Falsepositive}

\end{figure}

Figure~\ref{Falsepositive} shows the number of false positive per unit of time obtained by analyzing, with the algorithm described above, 628s seconds of signals before the helicase injection, compared to the number of stepping events per unit of time during helicase unwinding at 500 nM ATP.
Figure~\ref{Step size distribution} shows the distribution of measured step sizes of 190 stepping events of the helicase Upf1.

\subsection{Viscosity and axial noise in the vicinity of the surface}
\label{noisebrenner}
\begin{figure}[H]
\centering
\includegraphics[width=0.8\linewidth]{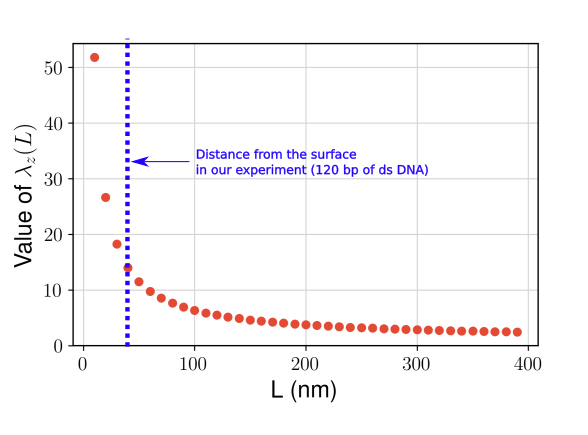} 

\caption{Evolution of the correction factor to Stoke's law $\lambda_z(L)$ for the drag of a sphere of radius 0.5 $\mu m$ when the sphere lies close to the surface as a function of the distance $L$ from the surface, according to the analytical formula of Brenner \cite{brenner1961slow}. In our experiment, the drag is 15 times larger than the one that would experience the bead infinitely far from the surface. }
\label{Brennercurve}

\end{figure}
\begin{figure}[H]
\centering
\includegraphics[width=0.8\linewidth]{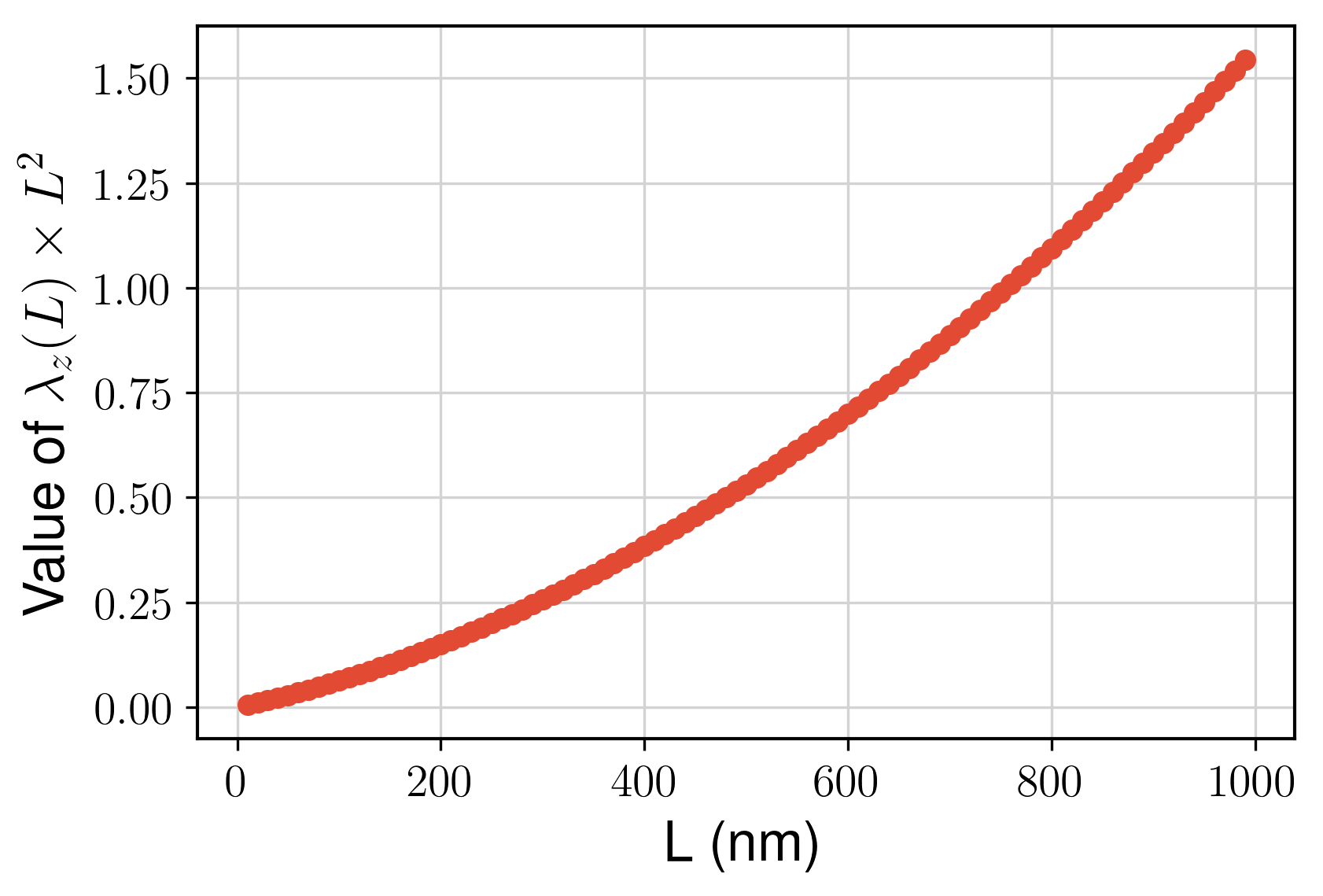} 

\caption{Evolution of the product  $\lambda_z(L) \times L^2$  as a function of L.}
\label{noiseL}

\end{figure}

\begin{figure}[H]

 \centering
 \includegraphics[width=0.8\linewidth]{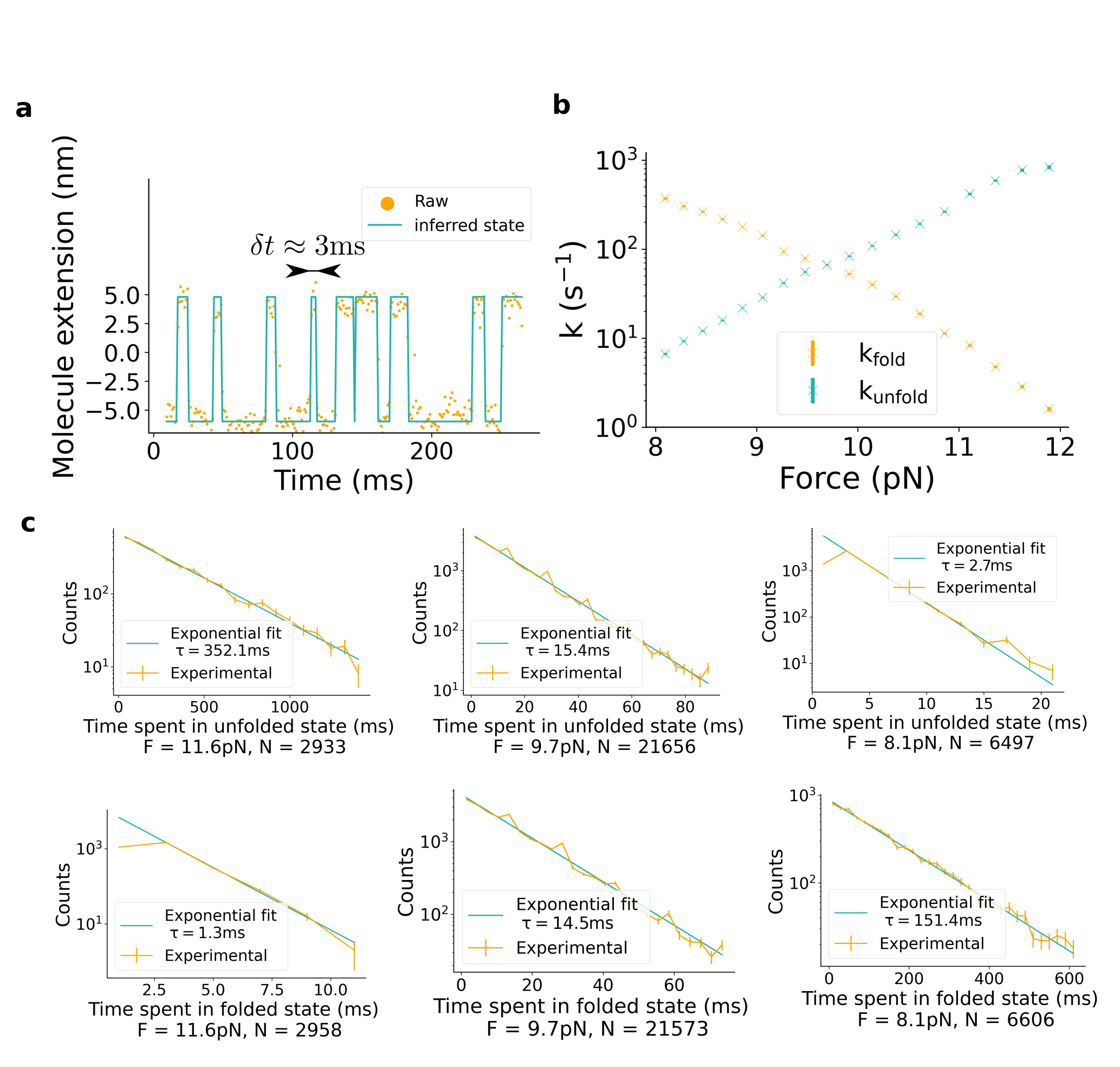}
 \caption{Kinetics of folding/unfolding of a 10 bp hairpin measured with \emph{SDI}. \textbf{a} : Trace representing the closing and the opening of a 10bp-hairpin at a acquisition frequency of 1500Hz. The high optical resolution of SDI allows a frame-by-frame determination of the state of the hairpin and thus  for a millisecond time resolution. \textbf{b} : Kinetics of opening/closing of the HP as a function of the applied force. There is an $\approx 10 \%$ uncertaintity on the force because of the variability of the bead magnetization. This uncertainty is not represented in the graph as it is systematic. \textbf{c} Distribution of folding/unfolding times at three different forces. Exponential fitting of the distribution allows for the determination of the kinetic parameters represented in b.}
 \label{HP10}
 \end{figure}

\begin{figure}[H]
\centering

\includegraphics[width=0.8\linewidth]{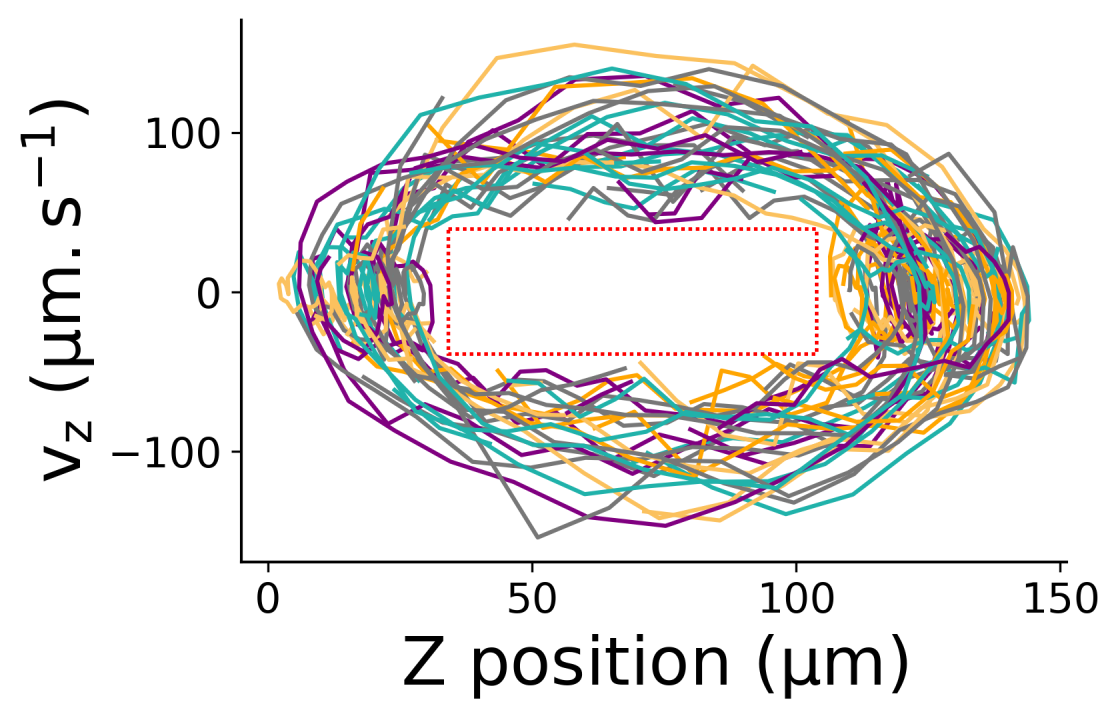} 
 \caption{Instantaneous velocity as a function of the vertical position of the algae of 97 trajectories are displayed. For the sake of clarity, among the 332 original single trajectories, we only displayed the ones that had no point in the excluded region represented by the red dashed rectangle. Represented trajectories are displayed in their entirety. }
\label{Vzvsz}
\end{figure}

 
In this section we briefly discuss the impact of the increased hydrodynamic drag close to the surface on the measurement noise. The noise density below the z-cutoff frequencies (in our case several kHz) for an overdamped Brownian sphere attached to a molecule of stiffness $k$ is written :
\begin{equation}
<P(0)> \propto \frac{k_B T \gamma}{k^2}  
\label{noiseplateau}
\end{equation}
The drag $\gamma$ thus plays an important role in the measurement noise. However, this drag deviates from Stoke's law in the vicinity of a surface. $\gamma(z) = 6\pi \eta R \lambda_z(L)$, where R is the radius of the bead and $\lambda$ has been computed analytically by Brenner (\cite{brenner1961slow}). By looking at the figure~\ref{Brennercurve}, this drag can be significantly reduced by working further away from the surface. This could be done by using larger dsDNA handles. However, the stiffness of such handles decreases with the number of monomers and thus with their lengths : $k \propto \frac{1}{N} \propto \frac{1}{L}$.
Thus, the dependence of the noise density as a function of the length of the handles verifies :
\begin{equation}
<P(0)> \propto \frac{k_B T \lambda_z(L)}{k(L)^2}  \propto L^2 \lambda_z(L)  
\label{noiseplateau2}
\end{equation}
We draw this function as a function of $L$ in figure~\ref{noiseL}. As can be seen, it is a growing function of $L$, thus showing that the loss of noise due to the decreased stiffness of the handles is worse than the gain in noise due to the decrease of the drag. For this reason, there is no hope that increasing the length of our dsDNA handles could help fighting against this surface effect. That is the reason why we propose in our discussion to use handles that are much stiffer than the DNA substrate whose change of extension we are measuring, in order to bring the bead further away from the surface without increasing significantly the stiffness of the construct. 

\subsection{Hairpin dynamics}
 
 \subsection{Algae dynamics}

\begin{figure}[H]
\centering
\includegraphics[width=0.8\linewidth]{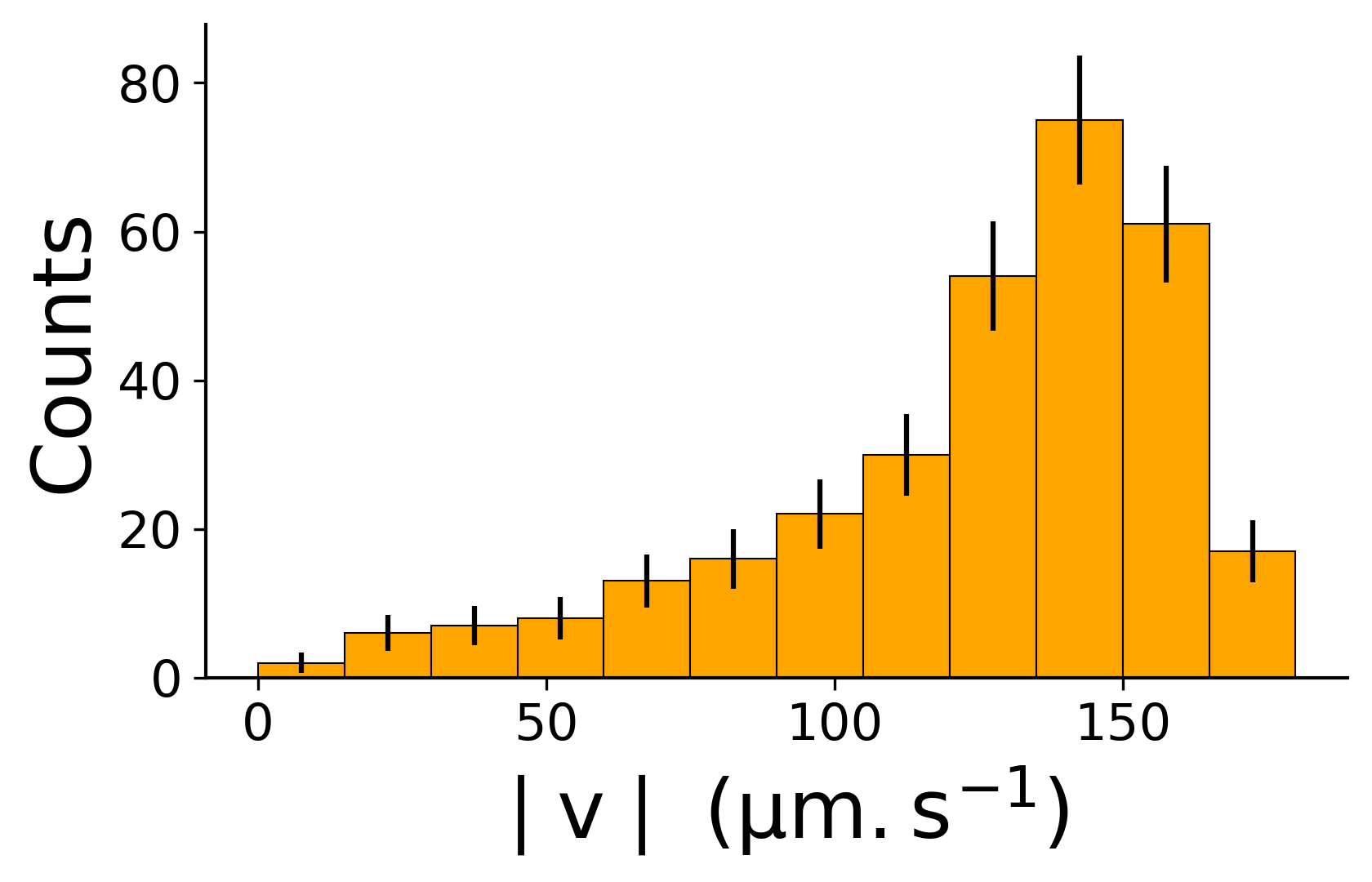} 

\caption{Mean velocity of the algae \emph{Chlamydomonas Reinhardtii} over 332 single trajectories. One count corresponds to one trajectory.}
\label{vdistrib}
\end{figure}

\subsection{Alignment procedure}
\label{alignsi}

We develop the following alignment method with the goal of positioning the amplitude mask 1) in the Fourier plane with a sub-millimetric accuracy and 2) centered on the optical axis of the system with the accuracy of ten microns.
As the Fourier plane is located inside the microscope objective, a 4f line is installed to provide a mechanical access on its image plane (pair of 1” achromatic doublet lens of 100mm, the focal length has been chosen to avoid 1)  vignetting effect and 2) off-axis aberrations of more powerful doublet lens). 

 \begin{figure}[H]
 \centering
 \includegraphics[width=0.8\linewidth]{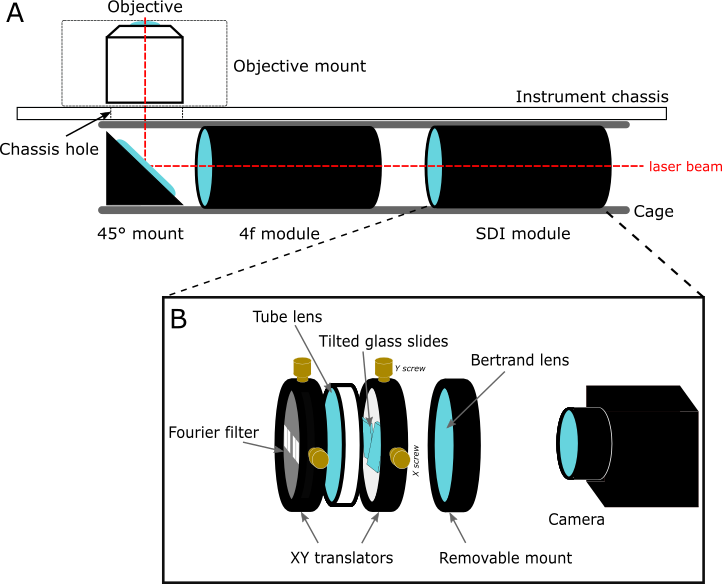}
 \caption{\textbf{A}. A folded 4f line (elliptic mirror + 2 100mm Thorlabs) is positioned below the objective to provide mechanical access to its Fourier plane.  \textbf{B}. The SDI module comprises 1) a compact assembly of amplitude mask-tube lens-tilted glass windows, 2) a removable Bertrand lens and 3) a camera.}
  \label{alignmain}
  \end{figure}

\textbf{Phase 1 : Assembly and Installation of the 4f line}

\begin{figure}[H]
 \centering
 \includegraphics[width=0.8\linewidth]{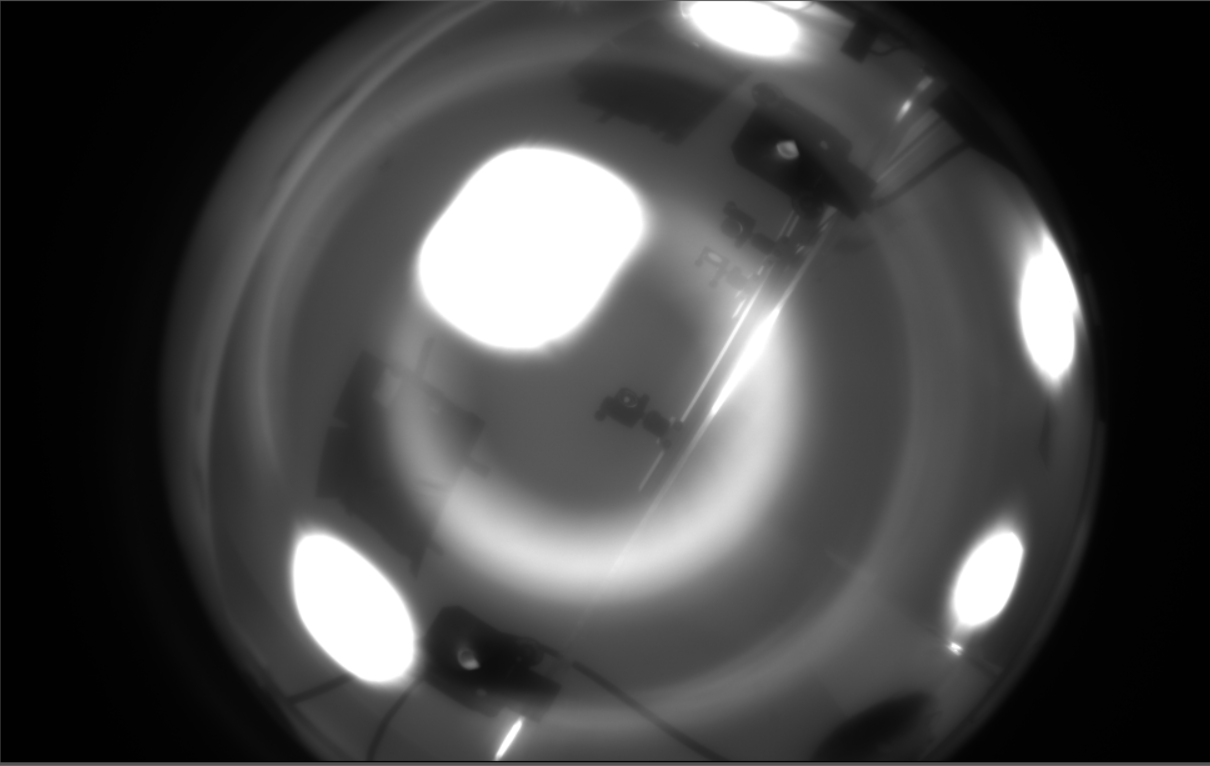}
 \caption{Image of the entrance pupil (telecentric objective) observed on a camera  temporarily placed at the focal plane of the second lens of the 4f line, after proper positioning of the 4f line in the system: the large black squares pieces of cardboard positioned on the ceiling of the room.}
  \label{align4f}
  \end{figure}

\begin{enumerate}
    \item The distance between the two lenses of the 4f line is locked (SM1 tubes from Thorlabs) once obtained sharp images of objects at the infinity. To do so, a temporary cage system assembles the 4f line and a module composed by a 50mm achromatic lens and a camera at its focal plane.
    \item The 4f line is installed in the cage system, attention is paid to align  it along the optical axis :
    \begin{enumerate}
        \item First, a temporary module composed by SM1 tubes and a camera are assembled such that once brought in contact with the 4f-module, the camera is placed at the focal plane of the second lens of the 4f-module.
        \item  The whole module (4f line + a temporary module) is then moved along the optical axis until one gets a sharp image of the pupil (see  Figure~\ref{align4f}).
    \end{enumerate}
\end{enumerate}

The assembly and integration of the SDI module itself employ a Bertrand lens, used to simultaneously obtain a sharp image of the mask containing the slits and of the back focal plane of the microscope objective (typically an object attached on the ceiling of the room).

\textbf{Phase 2: assembly of the SDI module.}

\begin{enumerate}

    \item 3 optical components (mask containing the slits - tube lens - pair of tilted glass slides, see Figure~\ref{alignmain}) are assembled in a cage system such that the tube lens is sandwiched tightly between the  mask and the tilted glass slides. 
    \item Once this module is assembled, the camera is added at the focal point of the tube lens (the tilted glass slides must be in place during this tuning).
    \item  A Bertrand lens (25mm of focal length, to perform 1:1 imaging between the place containing the slits and the camera plane) is added in the assembly, by the means of a magnetic mount (CPF90, Thorlabs), the position of which along the optical axis being tuned along the cage system axis until one gets a sharp image of the slits.
\end{enumerate}

 \begin{figure}[H]
 \centering
 \includegraphics[width=0.8\linewidth]{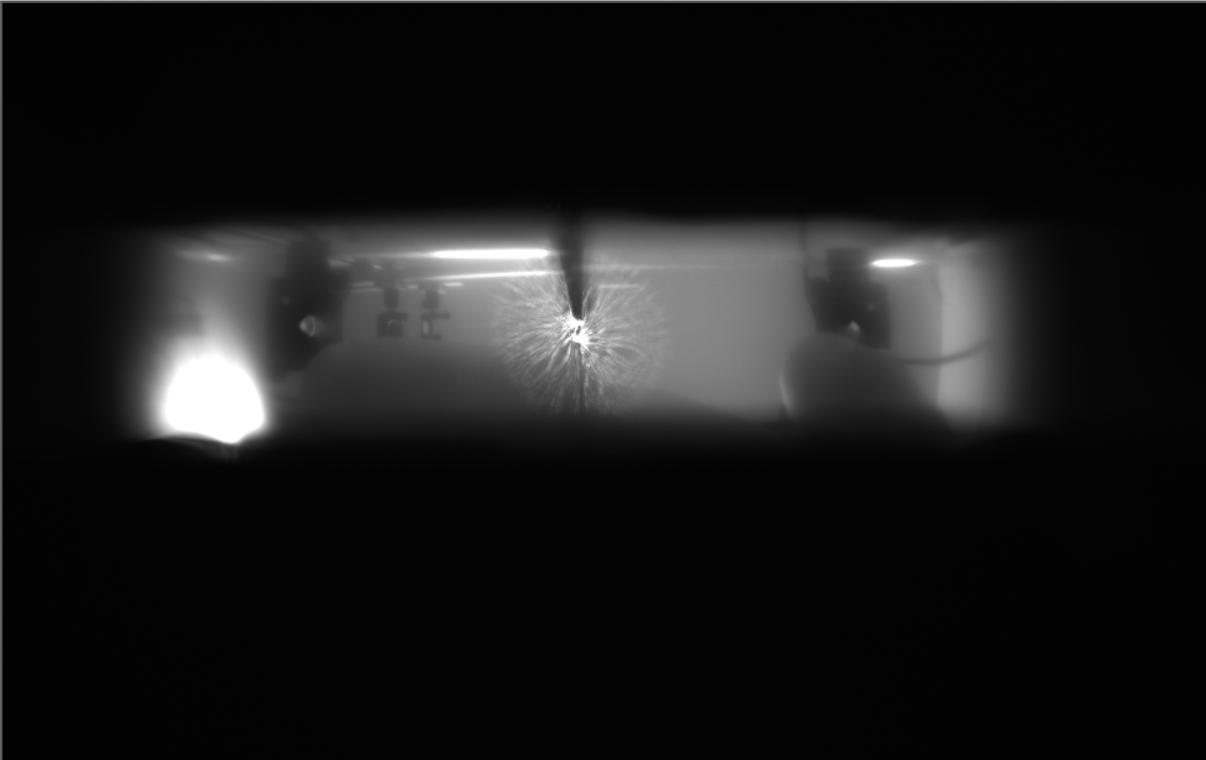}
 \caption{Image of the Fourier plane of the objective, here observed through the Bertrand lens of the SDI module. Slits have been removed, and the glass slits have been centered such that the spot of a HeNe alignment laser is shown split and largely cut by the line between the glass windows.}
  \label{alignB1}
  \end{figure}
  
   \begin{figure}[H]
 \centering
 \includegraphics[width=0.8\linewidth]{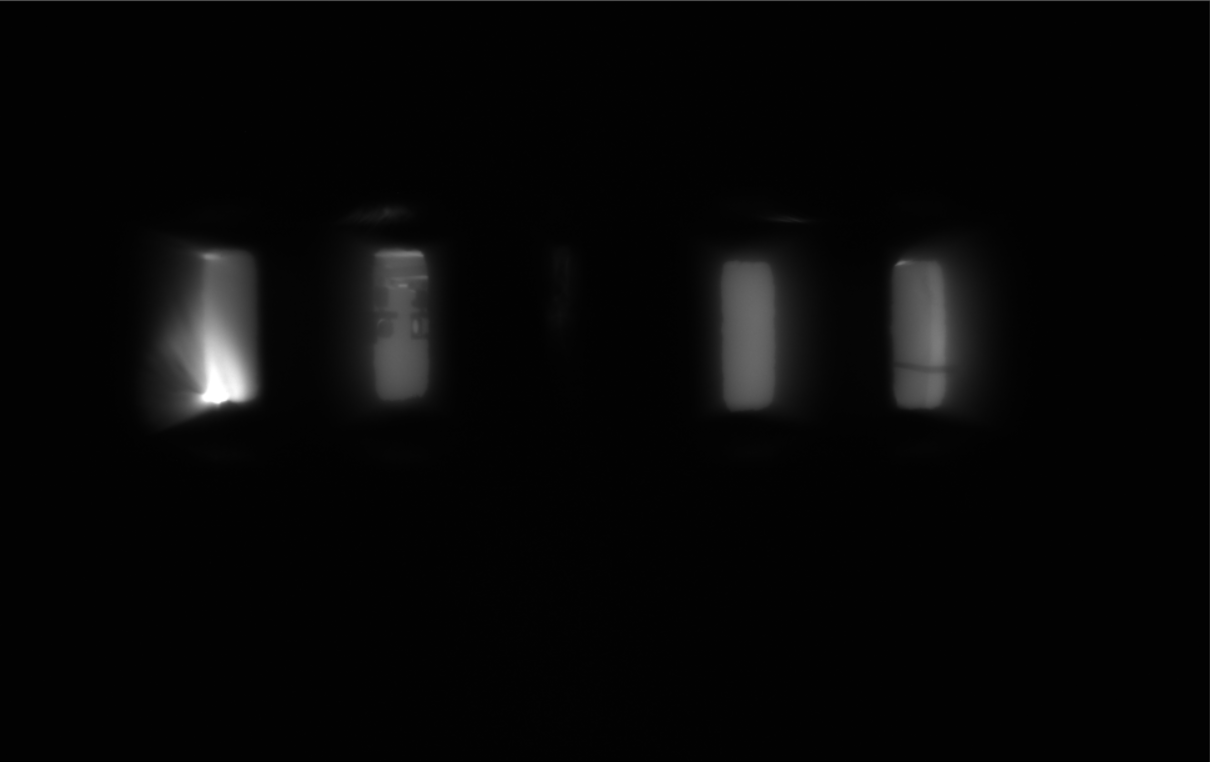}
 \caption{Same type of image, but with a mask added: slits of the mask and objects placed at an infinite distance of the objective are simultaneously sharp.}
  \label{alignB2}
  \end{figure}

\textbf{Phase 3: coarse implementation of the SDI module on the system}

\begin{enumerate}
    \item With Bertrand lens on position, the SDI module is inserted on the main cage system of the instrument and its position along the optical axis is manually tuned until one gets a sharp image of the back focal plane through the mask (Figure~\ref{alignD2}).
    \item The lateral position of the mask containing the slits and the glass slides are then adjusted to be centered onto the central of the pupil image (an alignment laser on top of the system can help).
    \item At that point, the illumination angles of the two light sources are optimized by ensuring that they converge in the center of each pair of slits.
\end{enumerate}

This protocol typically results in good quality SDI images. However : 
\begin{itemize}
    \item Residual error in the transverse positioning of the mask in the Fourier plane may result in the selection of slightly different angles for the left and right pairs of slits : this leads to a common differential sensitivity (variation of phase vs. defocus) between the two sets of fringes on the field of view. For example, all top fringes share the same offset of sensitivity compared to their corresponding bottom fringes.
    \item Residual error in the positioning of the mask along the optical axis (i.e., not truly in the Fourier plane) involves that the selected averaged angles may according to the position in the field of view. 
\end{itemize}

Consequently, a final alignment step is performed directly on the signals generated by fixed objects in the field of view of the microscope.

\textbf{Phase 4: fine adjustment of the SDI module.}

\begin{enumerate}
    \item 3 images obtained at focus are recorded, as well as images taken at large and symmetrical values of defocus (above the depth of field allowed by the setup, i.e. until the two optical rays for each fringe are separated).
    \item 3 objects are selected, located on the center and on both sides (along the mask axis) of the field of view.
    \item The transverse position of the mask is corrected (using micrometric screws), until on-axis image shows no differential sensitivity between both sets of fringes.
    \item Longitudinal position of the complete SDI module (sliding along the optical axis on the cage system) is corrected until one sees no different sensitivities between on-axis image and off-axis images (symmetric).
    \item Steps 1-3-4 are iterated until convergence.
\end{enumerate}
   
\begin{figure}[H]
 \centering
 \includegraphics[width=0.8\linewidth]{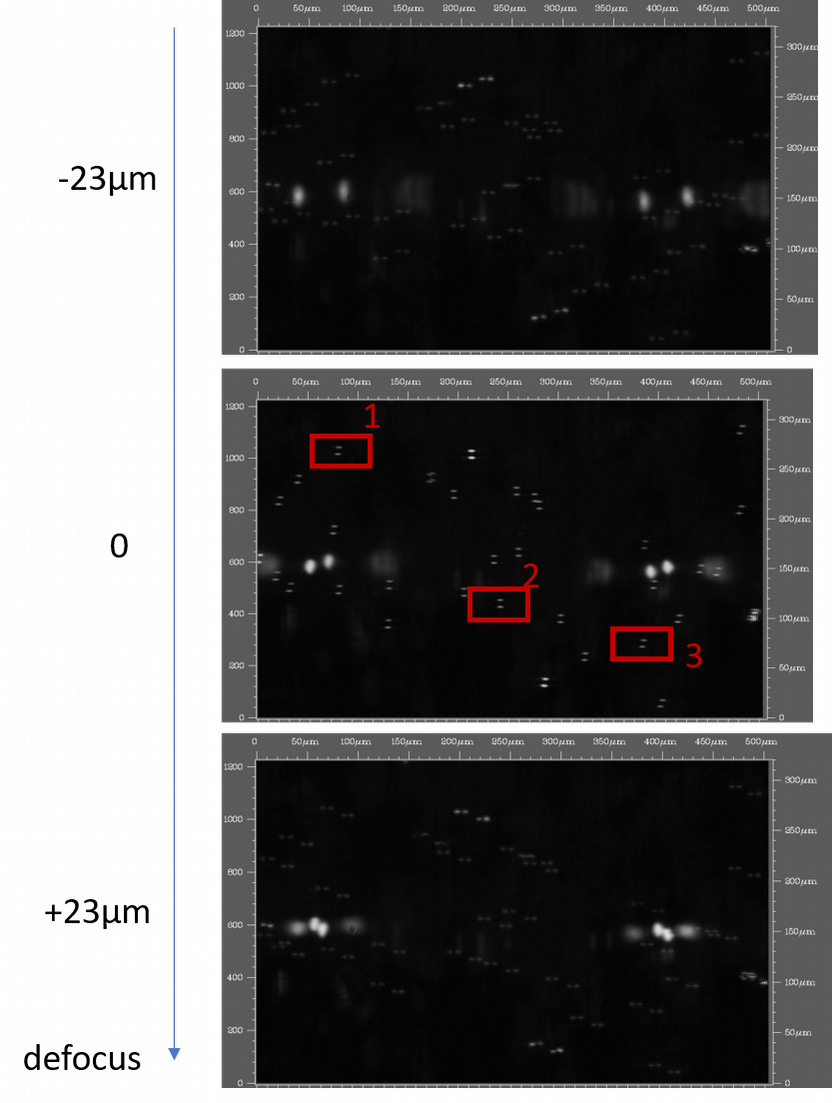}
 \caption{ Images recorded for -23,0,+23 $\mu m$ of defocus at the final position of the mask. Three objects are selected (see associated text). The large white spots are caused by the edges of a thermalization box present in the light path and are not related to the SDI imaging itself.}
  \label{alignD1}
  \end{figure}

\begin{figure}[H]
 \centering
 \includegraphics[width=0.8\linewidth]{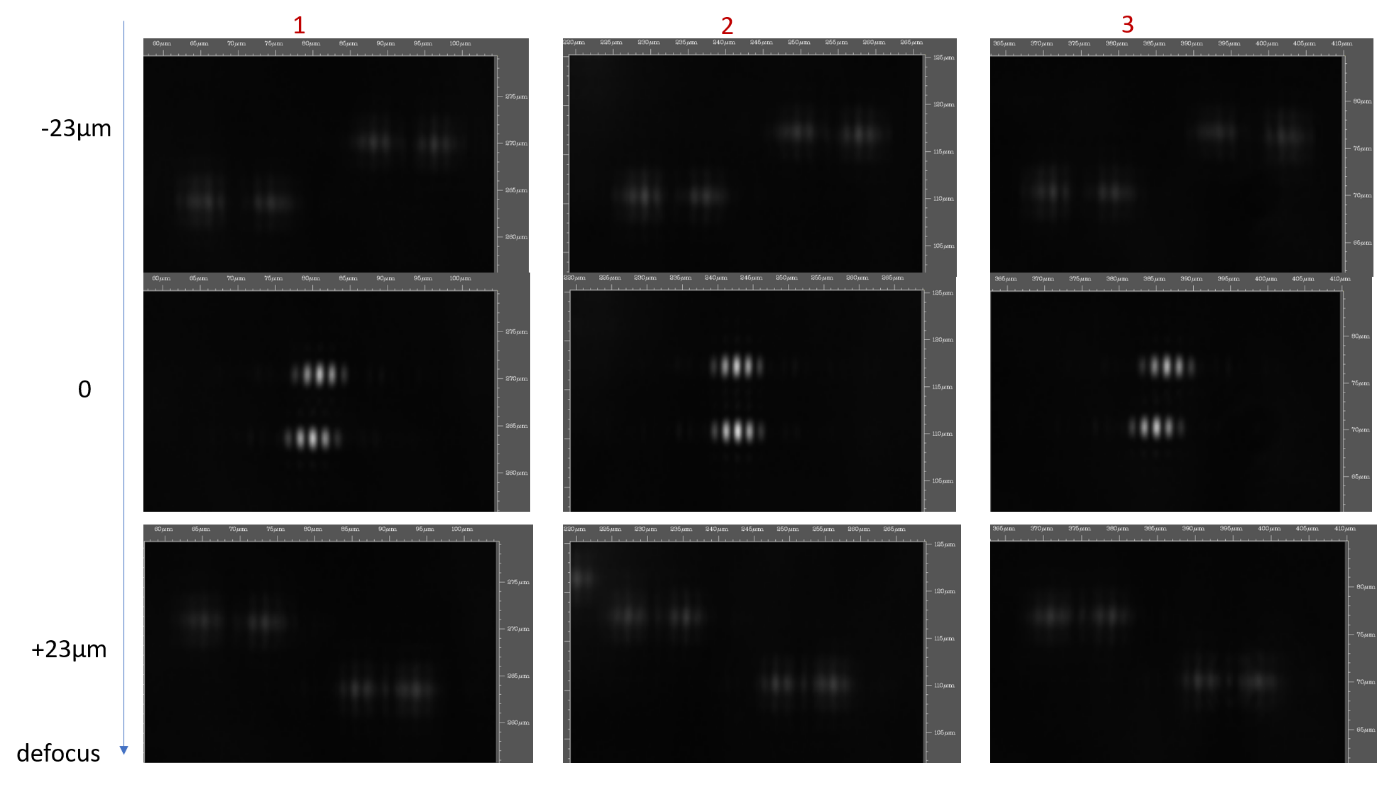}
 \caption{Images obtained for the 3 objects selected on the field of view and for the 3 focus positions.}
  \label{alignD2}
  \end{figure}

A typical set of images corresponding to one iteration of this process is shown on Figure~\ref{alignD1} and Figure~\ref{alignD2}. They have been recorded during the alignment of an SDI instrument featuring a 40X oil-immersion system. 

\begin{table}[H]
\resizebox{\textwidth}{!}{%

\begin{tabular}{|l|l|}
  \hline
  DNA name & Sequence \\
   (cf. Material and methods) & \\
  \hline
  Oli1 & {5' ATTCGAAGAGCACCAGAAAGACCAAAAGACACGGTGAAGGATTAGACAGAAGAAGAC 3'-DBCO} \\
  \hline
  Oli2  & 5' Double-biotin TGGGAGTAGCGGATCATGATGGATGTTGCCAGCTGGTATGGAAGCTAATAGCGCCGGT 3' \\
    \hline

  HP10STACK & 5' GTCTTCTTCTGTCTAATCCTTCACCGTGTCTTTTGGTCTTTCTGGTGCTCTTCGAATACTGCCAGAGTTTTCTCTGGC \\
   (stacking assay) & \ \ \ \ AGTGCGTGCTCGCAGTGACCGGCGCTATTAGCTTCCATACCAGCTGGCAACATCCATCATGATCCGCTACTCCCA 3' \\
     \hline

  Oliblock-half-stack & 5' ACTGCCAGAGAAAACT 3'\\
    \hline

  Oliblock-full-stack & 5' GCACGCACTGCCAGAGAAAACT 3'\\
\hline
 HP10FLUC & 5' GTCTTCTTCTGTCTAATCCTTCACCGTGTCTTTTGGTCTTTCTGGTGCTCTTCGAATTTTTTTTACTGCCAGAGTTT \\
  (fluctuation assay) &  TCTCTGGCAGTGCGTGCTCGCAGTGACCGGCGCTATTAGCTTCCATACCAGCTGGCAACATCCATCATGATCCGCTACTCCCA 3' \\
    \hline

  HP30 (UPF1) & 5' GTCTTCTTCTGTCTAATCCTTCACCGTGTCTTTTGGTCTTTCTGGTGCTCTTCGAATTTTTTTTAGTGCAGATGCTTCCTATAGACTGCCAGAGTTTTC \\
   & TCTGGCAGTCTATAGGAAGCATCTGCACTTTTTTTTACCGGCGCTATTAGCTTCCATACCAGCTGGCAACATCCATCATGATCCGCTACTCCCA 3' \\

  \hline
\end{tabular}}
\caption{DNA sequences used in this paper. Names are defined in the section Material and Methods.}
\label{dnaseq}
\end{table}

\subsection{Rotational postprocessing}

\begin{figure}[H]
    \centering
    \includegraphics[width=0.8\linewidth]{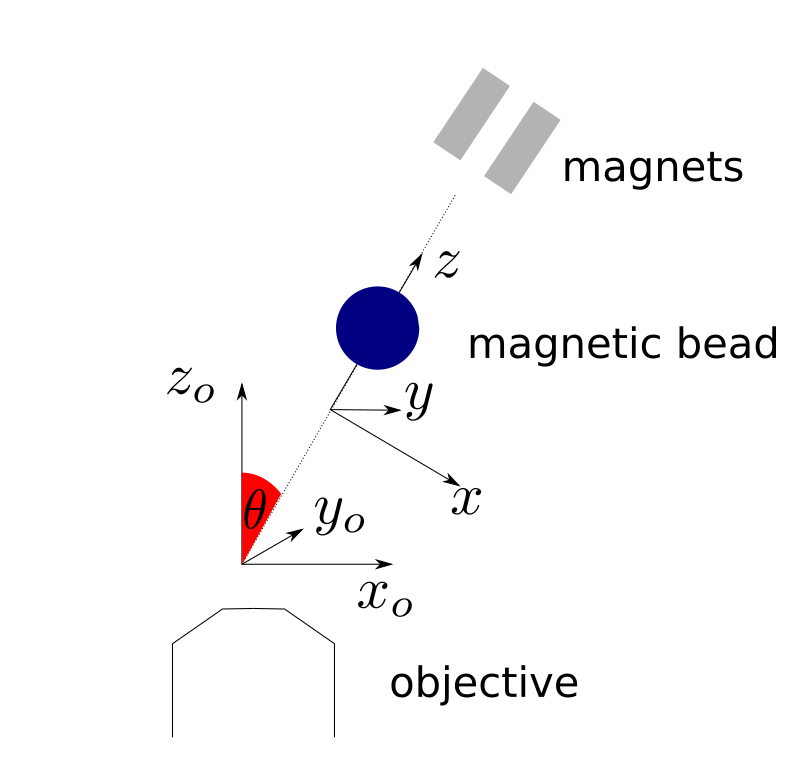}
    \caption{Schematic explanation of the notations for the following section. We claim that a slight misalignment between the magnetic axis, which is  relevant for the thermodynamic understanding of the Brownian noise, and the optical axis, explains part of the noise in magnetic tweezers experiments and can be easily corrected by a post-processing rotation. On this picture, the angle is exaggerated on purpose, but even an angle as small as $1^{\circ}$ can have a very important effect on the observed noise observed in the optical direction $z_o$. }
    \label{figurerot}
\end{figure}

A slight misalignment of the optical axis ($z_o$) with the axis of the magnetic force ($z$) can cause a large increase of the measured noise in the $z_o$ direction. 

Let's call $\theta$ the small angle between $z_o$ and $z$ and $u$ the coordinate of the bead position along the axis of rotation (that belongs to the plane (x,y)) perpendicular to $z$. For small $\theta$, $z_o$ verifies :

\begin{equation}
    z_o = z + \theta u
\end{equation}

and the fluctuations verify ($x$,$y$ and $z$ being uncorrelated at the first order approximation):

\begin{equation}
    \sigma_{z_o} = \sqrt{\sigma_z^2 +  \theta^2 \sigma_u^2}
    \label{bruitax}
\end{equation}
\begin{equation}
    \sigma_{z_o} \simeq \sigma_z ( 1 +   \frac{\theta^2}{2} \frac{\sigma_u^2}{\sigma_z^2})
    \label{bruitax2}
\end{equation}

We recall the formulation of the power spectrum density of an overdamped trapped bead :
\begin{equation}
<P(\omega)> \propto \frac{k_B T}{\gamma} \frac{1}{\omega ^2+ \omega_c ^2}
\end{equation},
where $\gamma$ is the viscous coefficient of the bead, $k$ the stiffness of the trap, and $\omega_c = \frac{k}{\gamma}$

For the frequencies below the cut-off frequencies, the noise density is roughly :
\begin{equation}
<P(\omega)> = <P(0)> \propto \frac{k_B T \gamma}{k^2}  
\label{noiseplateau3}
\end{equation}

For a polymer of length $L$ whose thermodynamics properties are encoded in its force-extension curve $F(L)$, and that is subject to a force $F$ in the $z$ direction;
\begin{equation}
k_{\parallel} =  k_z = \frac{\partial F}{\partial L}
 \end{equation}
 \begin{equation}
 k_{\bot} = k_x = k_y = \frac{F}{L}
\end{equation}

For a 120 bp double-stranded DNA subject to a force of $10 pN$, we have roughly :
\begin{equation}
k_z = \frac{\partial F}{\partial L} \simeq 10 pN/nm
 \end{equation}
 \begin{equation}
 k_x = k_y \simeq 0.25 pN/nm
\end{equation}

Thus, the noise density below the cut-off frequency is larger in the lateral direction by a factor $\frac{k_{\parallel}^2}{k_{\bot}^2} = 1600$, and thus the integrated noise at such an acquisition frequency  is bigger in the lateral direction ($\sigma_{\bot}$) compared to the axial direction ($\sigma_{\parallel}$) by a factor $\sqrt{1600} = 40$. 

For this reason, given eq.~\ref{bruitax}, an angle of $ 2 ^{\circ}$ between the optical axis and the magnetic axis is large enough to double the measured noise in the optical axial position $z_o$ compared to the noise in the magnetic direction $z$.

\begin{figure}[H]
\centering
\includegraphics[width=0.8\linewidth]{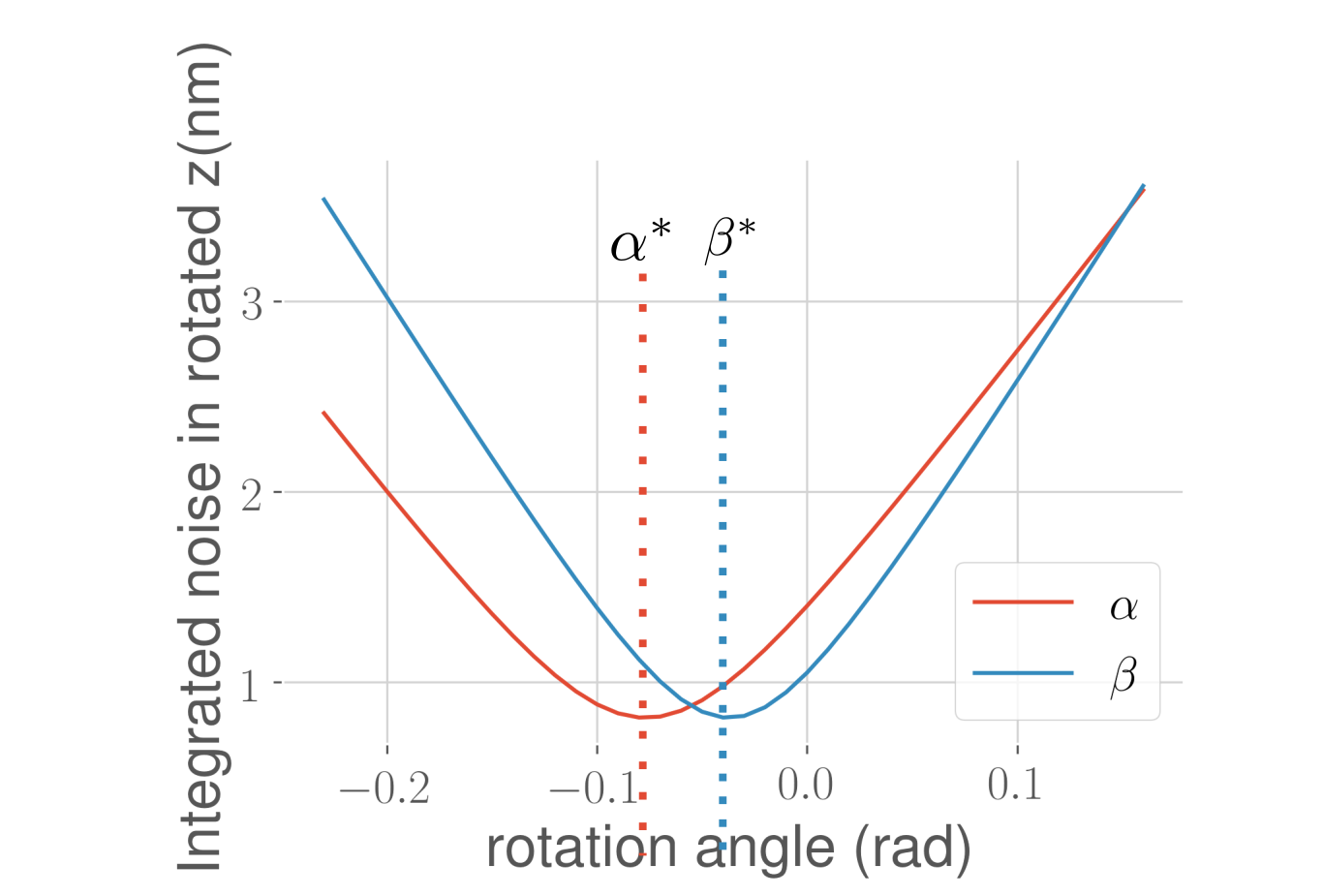} 
\caption{Measured noise in the z-direction obtained through rotation of the measured data ($x_o(t)$,$y_o(t)$,$z_o(t)$) with angles ($\alpha$,$\beta$) respectively around the x and the y axis. The curve as a function of $\alpha$ is drawn for $\beta=\beta^*$. The curve as a function of $\beta$ is drawn for $\alpha=\alpha^*$.}
\label{sigmavsalphabeta}
\end{figure}

\begin{figure}[H]
\centering
\includegraphics[width=0.8\linewidth]{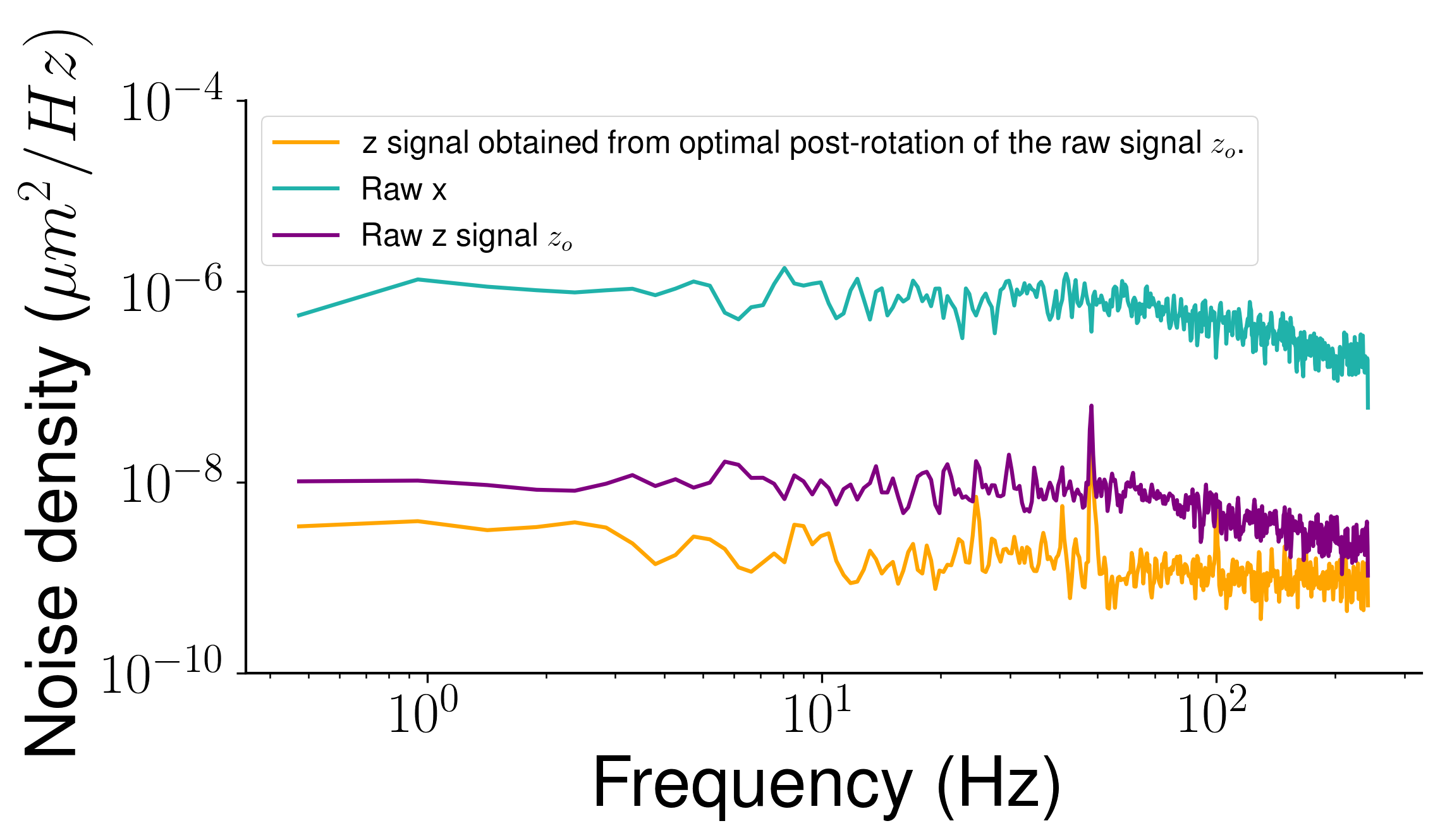} 

\caption{Power spectrum comparison. In turquoise, the power spectrum of $x_o(t)$. In purple, the power spectrum of $z_o(t)$. In orange, the power spectrum of $z(t)$ with $z$ being taken as the axis corresponding to the rotation that minimizes the noise in the $z$ direction. The cutoff frequency of the X-signal is visible in the $z_o(t)$ trace, demonstrating that this axis is coupled with the $x$ direction. The orange curve shows that rotating the signal allows decoupling the signals.}
\label{Spectrerotation}

\end{figure}

\begin{figure}[H]
\centering
\includegraphics[width=0.8\linewidth]{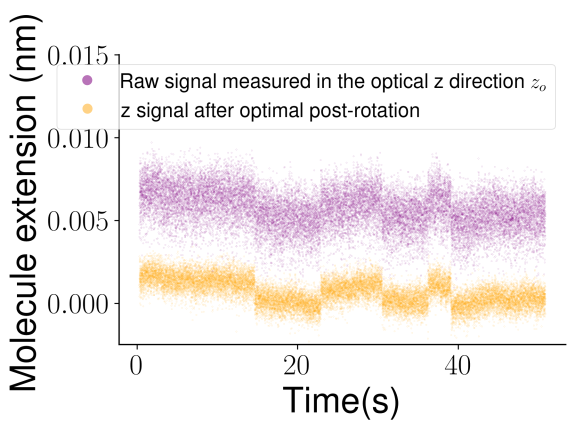} 

\caption{Traces obtained during the oligonucleotide hybridization assay described in the main text. In purple,  raw $z_o(t)$ . In orange, optimally rotated $z(t)$. The rotation allows to  better distinguish the 1 nm steps in the data. The curves are vertically offset for clarity.}
\label{oligorotation}

\end{figure}

To overcome this issue, we post-rotate the experimental data in order to recover the magnetic direction $z$ from the measurement in the orthogonal referential ($x_o$,$y_o$,$z_o$). We look for the 3D rotation angles ($\alpha,\beta$) that minimize the noise in the rotated $z$ direction. Formally, we minimize $\sigma_{z}$ with respect to $\alpha$ and $\beta$ where : 
\[
    \begin{bmatrix}
x \\
y  \\
z 
\end{bmatrix}
= R(\alpha,\beta)
    \begin{bmatrix}
x_o \\
y_o  \\
z_o 
\end{bmatrix}
\]
This minimization is obtained through standard multidimensional gradient-descent algorithms.
In figure~\ref{sigmavsalphabeta}, we show $\sigma_{z_o}$ as a function of $\alpha$ and $\beta$ and materialize the angles $\alpha^*$ and $\beta^*$ that minimizes the noise.
Figure~\ref{Spectrerotation} shows the effect of the post-processing rotation on the spectrum of a tethered bead. Figure~\ref{oligorotation} shows how the post-processing allows recovering signal of a much higher quality for the hybridization assay presented in the main text.  


\end{document}